%% file: Main_PASJ.tex
\documentclass{pasj01}

\Received{}
\Accepted{}
 
\usepackage[T1]{fontenc} 
\usepackage[switch,mathlines]{lineno}

\newcommand{\twelvecoh}{\mbox{$^{12}$CO($J$ = 2--1)}} 
 
\newcommand{\thirteencoh}{\mbox{$^{13}$CO($J$ = 2--1)}}

\newcommand{\msun}{\mbox{$M_\odot$}}
\newcommand{\kms}{\mbox{km~s$^{-1}$}}
\newcommand{\kkms}{\mbox{K~km~s$^{-1}$}}
\newcommand{\vlsr}{\mbox{$V_\mathrm{LSR}$}}

\newcommand{\hone}{\mbox{H{\sc i}}}
\newcommand{\htwo}{\mbox{H{\sc ii}}}

\usepackage{natbib}
\usepackage{enumerate}
\begin{document} 

\title{ 
A Kinematic Analysis of the Giant Molecular Complex W3; 
Possible Evidence for Cloud-Cloud Collisions that Triggered OB Star Clusters in W3 Main and W3(OH)}

\author{Rin I. \textsc{Yamada}\altaffilmark{1}}
\altaffiltext{1}{Department of Physics, Nagoya University, Furo-cho, Chikusa-ku, Nagoya 464-8601, Japan}
\email{yamada@a.phys.nagoya-u.ac.jp}

\author{Hidetoshi \textsc{Sano}\altaffilmark{1, 2}}
\altaffiltext{2}{Faculty of Engineering, Gifu University, 1-1 Yanagido, Gifu 501-1193, Japan}

\author{Kengo \textsc{Tachihara}\altaffilmark{1}}

\author{Rei \textsc{Enokiya}\altaffilmark{3}}
\altaffiltext{3}{Faculty of Science and Engineering, Kyushu Sangyo University, 2-3-1 
Matsukadai, Fukuoka 813-8503, Japan}

\author{Atsushi \textsc{Nishimura}\altaffilmark{4}}
\altaffiltext{4}{Nobeyama Radio Observatory, National Astronomical Observatory of Japan (NAOJ), National Institutes of Natural Sciences (NINS), 462-2 Nobeyama, Minamimaki, Minamisaku, Nagano 384-1305, Japan}

\author{Shinji \textsc{Fujita}\altaffilmark{5}}
\altaffiltext{5}{Institute of Statistical Mathematics, 10-3 Midori-cho, Tachikawa, Tokyo, Japan}

\author{Mikito \textsc{Kohno}\altaffilmark{6}}
\altaffiltext{6}{Curatorial Division, Nagoya City Science Museum, 2-17-1 Sakae, Naka-ku, Nagoya, Aichi 460-0008, Japan}

\author{John H. \textsc{Bieging}\altaffilmark{7}}
\altaffiltext{7}{Steward Observatory, The University of Arizona, Tucson, AZ 85721, USA}

\author{Yasuo \textsc{Fukui}\altaffilmark{1}}



\maketitle
\input{abst}

\input{intro}
\input{datasets}
\input{results}

\input{discussion}
\input{conc}

\input{ack}
\bibliography{reference}
\input{appendix}
\newpage

\end{document}

%% file: abst.tex
\begin{abstract}
W3 is one of the most outstanding regions of high-mass star formation in the outer solar circle, including two active star-forming clouds, W3 Main and W3(OH). Based on a new analysis of the $\twelvecoh$ data obtained at 38$^{\prime\prime}$ resolution, we have found three clouds having molecular mass from 2000 to 8000~$\msun$ at velocities, $-50$~$\kms$, $-43$~$\kms$, and $-39$~$\kms$. The $-43$~$\kms$ cloud is the most massive one, overlapping with the $-39$~$\kms$ cloud and the $-50$~$\kms$ cloud toward W3 Main and W3(OH), respectively. In W3 Main and W3(OH), we have found typical signatures of a cloud-cloud collision, i.e., the complementary distribution with/without a displacement between the two clouds and/or a V-shape in the position-velocity diagram. We frame a hypothesis that a cloud-cloud collision triggered the high-mass star formation in each region. The collision in W3 Main involves the $-39$~$\kms$ cloud and the $-43$~$\kms$ cloud. The collision likely produced a cavity in the $-43$~$\kms$ cloud having a size similar to the $-39$~$\kms$ cloud and triggered the formation of young high-mass stars in IC~1795 2 Myr ago. We suggest that the $-39$~$\kms$ cloud is still triggering the high-mass objects younger than 1 Myr embedded in W3 Main currently. On the other hand, another collision between the $-50$~$\kms$ cloud and the $-43$~$\kms$ cloud likely formed the heavily embedded objects in W3(OH) within $\sim$0.5 Myr ago. The present results favour an idea that cloud-cloud collisions are common phenomena not only in the inner solar circle but also in the outer solar circle, where the number of reported cloud-cloud collisions is yet limited (Fukui et al. 2021, PASJ, 73, S1).
\end{abstract}

%% file: intro.tex
\section{Introduction}
\subsection{High-mass star formation}
High-mass stars are important in our understanding of galaxy evolution because their influence on the interstellar medium is substantial via ultraviolet radiations, stellar winds, and supernova explosions. It was argued in the literature that the interstellar medium can form low to high-mass stars under the combined effects of turbulence, magnetic field, and gravity (for a review, see \citealp{McKee2007}), whereas it was not directly shown that these effects alone can realize the formation of high-mass stars and massive clusters. It was also argued that the formation of high-mass stars may require some extra mechanism different from or additional to the mechanism for low-mass star formation \citep{Zinnecker}. The initial gas condition for star formation, for instance, may be a critical factor in order to realize the formation of high-mass stars. Numerical simulations have shown that the competitive accretion model \citep{Bonnell} or the monolithic collapse model \citep{Krumholz_mass_acc} are possible mechanisms, whereas they are not readily compared with the observations. This is because the key signatures for comparison with observations are not always clearly pinpointed by the theoretical studies, allowing large room in the interpretation of observations.

In the competitive accretion model, the initial condition is a massive cloud-star system of more than a few 1000~$\msun$ where individual stars compete to acquire mass gravitationally, whereas the model fails to explain the formation of an isolated small-mass system including isolated one or a few O stars which are distributed widely in the Galaxy (for a review see e.g., \citealp{Fukui_rev, Ascenso}). The monolithic collapse model assumes an initial condition including ``a massive gas cloud of 100~$\msun$ within a radius of 0.1~pc'', and it was numerically demonstrated that high-mass stars of 40 and 30~$\msun$ in a binary are formed (Krumholz et al. 2009). However, it remains an open question how such massive, dense initial clouds are prepared in interstellar space. The dense core has to be rapidly formed before the core is consumed by low-mass star formation. It has also been discussed in the literature that high-mass stars are formed in massive hot cores or in infrared dark clouds having very high column density (e.g., \citealt{Egan}). It is, however, a puzzle that they are not always associated with $\htwo$ regions, which must be formed by O/early B stars, raising a doubt if they are real precursors of high-mass star formation \citep{Tan_2014}. See also \citet{Motte_annual} for a review of relevant observations.

Another possibility to resolve the issue is external triggers to compress gas and induce star formation, which is called the triggered star formation. A model of such a trigger is the gas compression by expanding $\htwo$ regions forming next-generation subgroups of OB associations, which is called the sequential star formation \citep{Elmegreen_Lada}. According to the model, once high-mass stars are formed by a certain (unknown) mechanism, the stars can compress the surrounding gas by the stellar feedback to a gravitationally unstable gas layer, where the next-generation high-mass stars are formed. The trigger continues to form high-mass stars until the gas is exhausted, and the model may explain the age sequence of subgroups in OB associations such as observed in Orion \citep{Blaauw1964}. Another triggering mechanism is a cloud-cloud collision (CCC) between two clouds having supersonic velocity separation. In this scheme, the gas is compressed in the interface layer between the clouds to a gravitationally unstable state, leading to the formation of high-mass stars. This mechanism explains the formation of the high-mass stars in an initial state without high-mass stars and also explains the distribution and age of all the high-mass stars involved as caused by the configuration of the two clouds and is distinguished from the sequential star formation, which assumes the first generation high-mass stars.

CCCs were proposed in 1950--1960 \citep{Oort_1954, Oort_spitzer}. Based on the optical spectroscopy of stellar absorption lines, \citet{Oort_1954} suggested that collisions between the interstellar clouds take place at a supersonic velocity of 10--15~$\kms$. In NGC~1333, \citet{Loren} observed two CO components of different velocities separated by 2~$\kms$ and proposed that a collision between two CO components triggered the star formation in NGC~1333. Numerical simulations of a CCC by \citet{Habe_Ohta} demonstrated that the collision between two clouds of different sizes plays a role in efficiently compressing the gas to trigger star formation, which was supported by follow-up simulations by \citet{Anathpindika} and \citet{Takahira2014}. The simulations were refined by incorporating the magnetic field by \citet{Inoue_Fukui} and \citet{Inoue_2018}, which showed that collisions are an effective process to trigger the formation of massive dense cores, a precursor of the high-mass star(s). Recently, \citet{Fukui_core_mass} compared these theoretical results with observations of CCCs and showed that the top-heavy mass spectra of the dense cores are consistent with observations. \citet{Fukui_rev} summarised the observational and theoretical understandings of star formation triggered by CCCs based on more than 50 candidate objects, which covers a wide range of the stellar mass from a single O star to massive clusters of $10^{6}$~$\msun$ discovered by extensive CO surveys of our Galaxy and the Local Group members. Consequently, recent studies favour CCCs as an important mechanism of high-mass star formation rather than the turbulent clouds without external triggers.

\subsection{W3 Giant Molecular Complex}
W3 is a radio continuum source discovered by \citet{Westerhout1958} and is a $\htwo$ region complex associated with a giant molecular complex (GMC) (for a review, see \citealp{Megeath2008}). Located in the Perseus Arm at a distance of 2~kpc \citep{Navarete2019}, W3 is the most active site of star formation in the Perseus Arm \citep{Ogura_Ishida, Oey_2005, Navarete2011, Navarete2019,Kiminki2015}, showing exceptionally active star formation in the outer solar circle. W3 is continuous to another $\htwo$ region named W4 corresponding to Perseus/Chimney superbubble \citep{Dennison} along the Galactic Plane. In the boundary between W3 and W4, there is a high-density layer (HDL), which is a molecular layer nearly vertical to the Galactic plane \citep{Lada_w3}. The HDL harbours several star-forming regions such as W3 Main, W3(OH), and AFGL~333, as well as diffuse $\htwo$ region IC 1795 and more compact, bright $\htwo$ region NGC~896. Each of these regions is associated with young stellar clusters with ages younger than 5 Myr, indicating recent star-forming activity.

Stellar clusters in the HDL have an age difference of a few Myr. The stellar cluster in IC~1795 is not associated with molecular gas but is associated with ionised gas, whereas OB stars in W3 Main and W3(OH) are deeply embedded in dense molecular material. According to the spectroscopic observations \citep{Oey_2005}, OB stars in IC 1795 have ages of 3--5~Myr while the age determination has uncertainties due to the assumption of stellar evolution models and extinction law. The primary source of the ionisation in IC~1795 is BD+61~411, an O6.5V star \citep{Oey_2005}. W3 Main is associated with the most massive cluster in W3, which has a total stellar mass of 4000~$\msun$ according to the near-infrared observations \citep{Bik2012}. The members of the cluster are approximately 10 OB stars, hyper-compact and ultra-compact $\htwo$ regions, as well as cold prestellar cores, which are localised within a diameter of $\sim$3~pc \citep{Claussen1994, Tieftrunk1997, Bik2012, Mottram2020}. In particular, an infrared source IRS~5 in W3 Main is known to harbour a Trapezium-like cluster \citep{Abt2000, Megeath2005, Rondon2008}. In the W3(OH) region, OH masers and H$_2$O masers are localised within less than 0.1~pc \citep{Forster, Reid1980}. The OH maser is excited by an O9 star \citep{Hirsch2012} and is associated with an ultra-compact $\htwo$ region with a diameter of  0.012~pc. Further, five early-type stars of B0--B3 are distributed in the W3(OH) region \citep{Navarete2011, Bik2012, Kiminki2015, Navarete2019}.

The sequential star formation model \citep{Elmegreen_Lada} has been discussed in W3 as a large-scale process of star formation. 
\citet{Lada_w3} made a large-scale CO survey of the W3/W4/W5 region at 8$^\prime$ resolution covering over 50~pc in the $^{12}$CO($J$~=~1--0) emission and proposed the sequential star formation. These authors discovered and mapped the HDL and found that the HDL is forming stars younger than those in W4 by $\sim$10~Myrs, and suggested that the next generation stars were formed by the trigger of the expansion of W4. \citet{Oey_2005} performed UBV photometry and derived the age distribution of the stars in IC~1795. These authors proposed that W4 (stellar age $\sim$20~Myr) triggered the star formation in IC~1795 (stellar age $\sim$3--5~Myr), and IC~1795 triggered the star formation in W3 Main and W3(OH). The authors named the whole process ``Hierarchical triggering''.

Recent higher-resolution observations at multi-wavelength, however, indicate that these simple pictures of sequential star formation may not work from distributions of stars and stellar ages. \citet{Feigelson} extensively studied stars of the low-mass populations with the Chandra X-ray satellite in W3 Main, IC~1795, and W3(OH). As a result, the authors agreed that stars in W3(OH) might have been triggered by IC~1795. On the other hand, the cluster of young stellar objects (YSOs) in W3 Main shows a large, spherical, centrally-condensed distribution, which was not interpreted as a consequence of the trigger by IC~1795.

\citet{Bik2012, Bik2014} made infrared photometry by the LBT (Large binocular telescope) Near Infrared Spectroscopic Utility with Camera and Integral Field Unit for Extragalactic Research (LUCI). As a result, these authors suggested that star formation in W3 Main began 2--3~Myr ago, and is continuing until now, causing a dispersion in age of a few Myr. It was also shown that the youngest stars lie in the central part of the cluster. Further, these authors suggested that feedback from the older stars in the W3 Main region does not propagate deeply into dense clouds, by more than 0.5~pc, probably due to higher initial gas density even after a few Myr after the cluster formation, and does not influence the cluster formation as a whole.

\citet{Roman_Zuniga} carried out an extensive study of the YSO age distribution through a new $JHK$ imaging by using the Calar Alto Observatory 3.5~m telescope combined with Chandra Source Catalog, mosaics of SPIRE and PACS on board \textit{Herschel}, and Bolocam 1.1~mm mosaic observations. These authors showed that there are many Class II sources in IC~1795 in spite of that the molecular gas is largely dispersed already, and suggested that star formation has been very active until about 2~Myr ago. In addition, it was shown that W3 Main and W3(OH) include a stellar population with an age of more than 2~Myr. Based on these, the authors concluded that it is hard to explain that the star formation in W3 Main and W3(OH) was triggered by stellar feedback from IC~1795 because of no significant age difference among the ages of IC~1795, W3 Main, and W3(OH).

These previous works presented doubts about the sequential star formation model, whereas an alternative star formation scenario based on CO observation and analysis of gas kinematics is not provided. It, therefore, still remains an open issue how star formation took place in IC~1795, W3 Main, and W3(OH).

\subsection{The aim of the present paper}
Star formation in W3 has not been fully understood yet in part due to the lack of a 
comprehensive study of the detailed gas kinematics. Recent studies extensively revealed 
the stellar properties in the region, which include both high-mass and low-mass stars at 
sub-mm and X-rays, as well as the deeply embedded stars. However, molecular 
gas kinematics has not been investigated deeply in spite of the high resolution and 
wide-field mapping of CO, and the other tracers have become available in the last two decades. 
We therefore commenced a systematic study of the molecular gas obtained by \cite{Bieging2011} with the Heinrich Hertz Sub-millimeter Telescope (HHT), which 
is employed using the analysis tools of gas kinematics developed in the last decade by the Nagoya group on the molecular clouds \citep{Fukui_rev}: e.g., velocity channel distribution, position-velocity diagram, separation of individual clouds. In the study, we aim to reveal a star formation mechanism that is dominant in W3 by utilizing the gas kinematic details and exploring its implications for high-mass star formation. Since the AFGL~333 region was investigated by \cite{Nakano} and \cite{Liang2021}, we will deal with the region in the present paper.

This paper is organized as follows; Section \ref{sec:data} describes the datasets employed and Section \ref{sec:results} gives the results of the present kinematical analysis. In Section \ref{sec:discussion} we discuss the high-mass star formation in IC~1795, W3~Main, and W3(OH), and conclude the paper in Section \ref{sec:conclusion}. We use the Galactic coordinate to point directions in which, for example, ``north'' means ``Galactic north'' in this paper. We refer to the Local Standard of Rest velocity ($\vlsr$) in the entire article.

\begin{figure}
	\includegraphics[width=8.5cm]{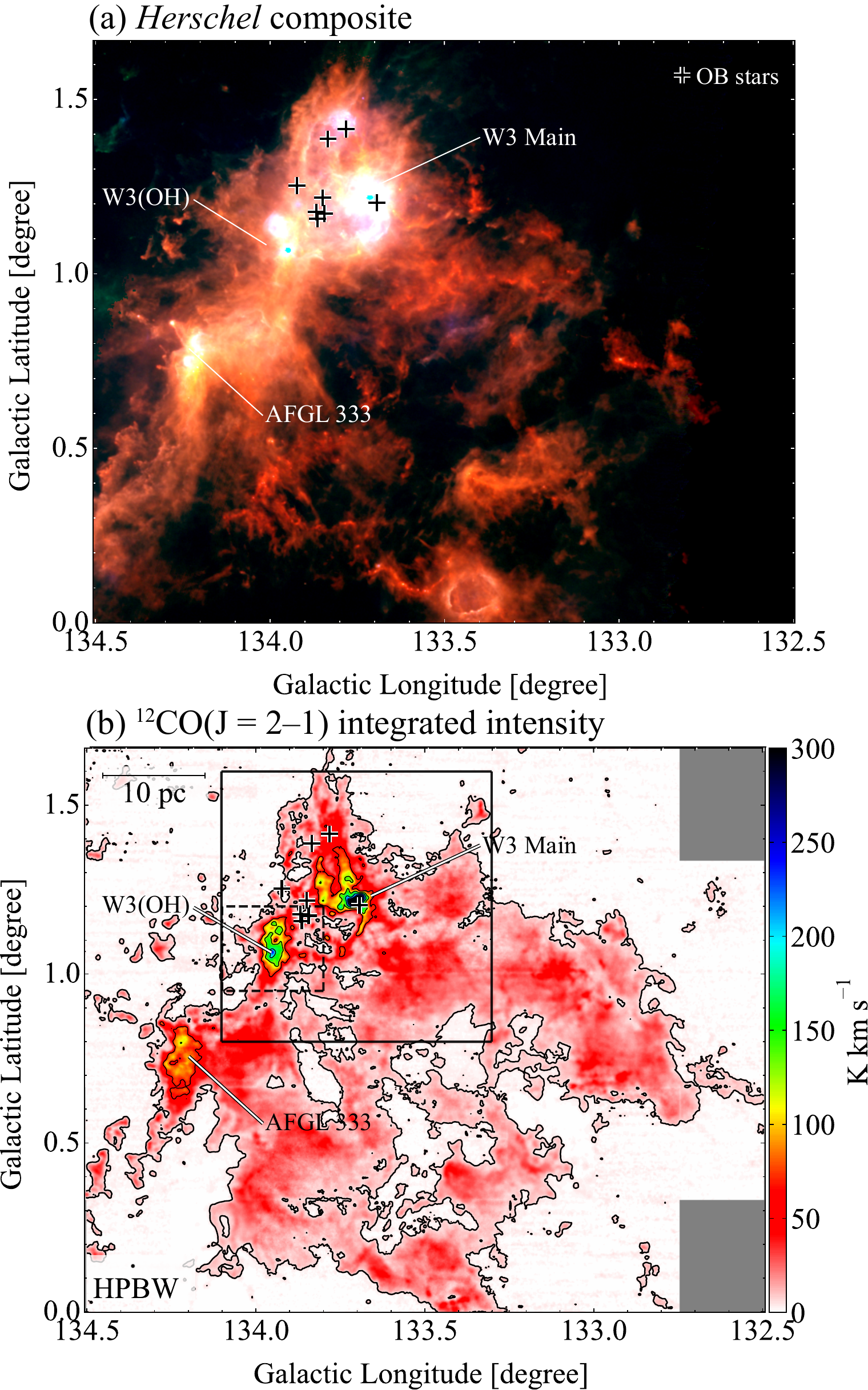}
    \caption{(a) Three-color composite image of the W3 GMC using PACS~70$~\mu$m (Blue), 160~$\mu$m (Green), and SPIRE~250~$\mu$m (Red) on board {\it{Herschel}}. The colour tables have been manipulated to bring out the structural detail in the map.  (b) Integrated intensity distribution of the $\twelvecoh$ emission for 
    a velocity range from $-53$ to $-28$~$\kms$. The black boxes shown by the solid 
    and dotted lines indicate the regions to be appeared in Figures \ref{fig4} and \ref{fig13}, 
    respectively. Two rectangular regions filled in grey are not covered in the present data. The black crosses represent the positions of O-type stars detected 
    by \citet{Oey_2005}. The lowest contour level and contour intervals correspond 
    to 3 and 60~$\kkms$, respectively.}
    \label{fig1}
\end{figure}

%% file: datasets.tex
\section{Datasets}\label{sec:data}
We used the $J$ = 2--1 line of $^{12}$CO and $^{13}$CO archived data observed with HHT at the Arizona radio observatory \citep{Bieging2011}. Observations were carried out in June, 
2005 to April, 2008. They used the On-The-Fly (OTF) method to map a region of 2\fdg00 times 
1\fdg67 square degrees. The map center was ($l$, $b$) = (133\fdg50, 0\fdg835). All the data were convolved to a spatial resolution of 38$\arcsec$. The velocity resolution is 1.30~$\kms$ at $^{12}$CO and 1.36~$\kms$ at $^{13}$CO, while they chose a common sample velocity of 0.5~$\kms$. The typical root-mean-square (RMS) noise temperatures of the $^{12}$CO emission and $^{13}$CO emission are 0.12~K and 0.14~K, respectively.
We prefer to use $^{12}$CO mainly instead of $^{13}$CO in the present work because $^{12}$CO can better trace the low density extended molecular gas. This is crucial in identifying the cloud interaction, while it may be saturated in small dense regions.

%% file: results.tex
\section{Results}\label{sec:results}
\subsection{The overall distributions of the molecular gas}

\begin{figure*}
	\includegraphics[width=17cm]{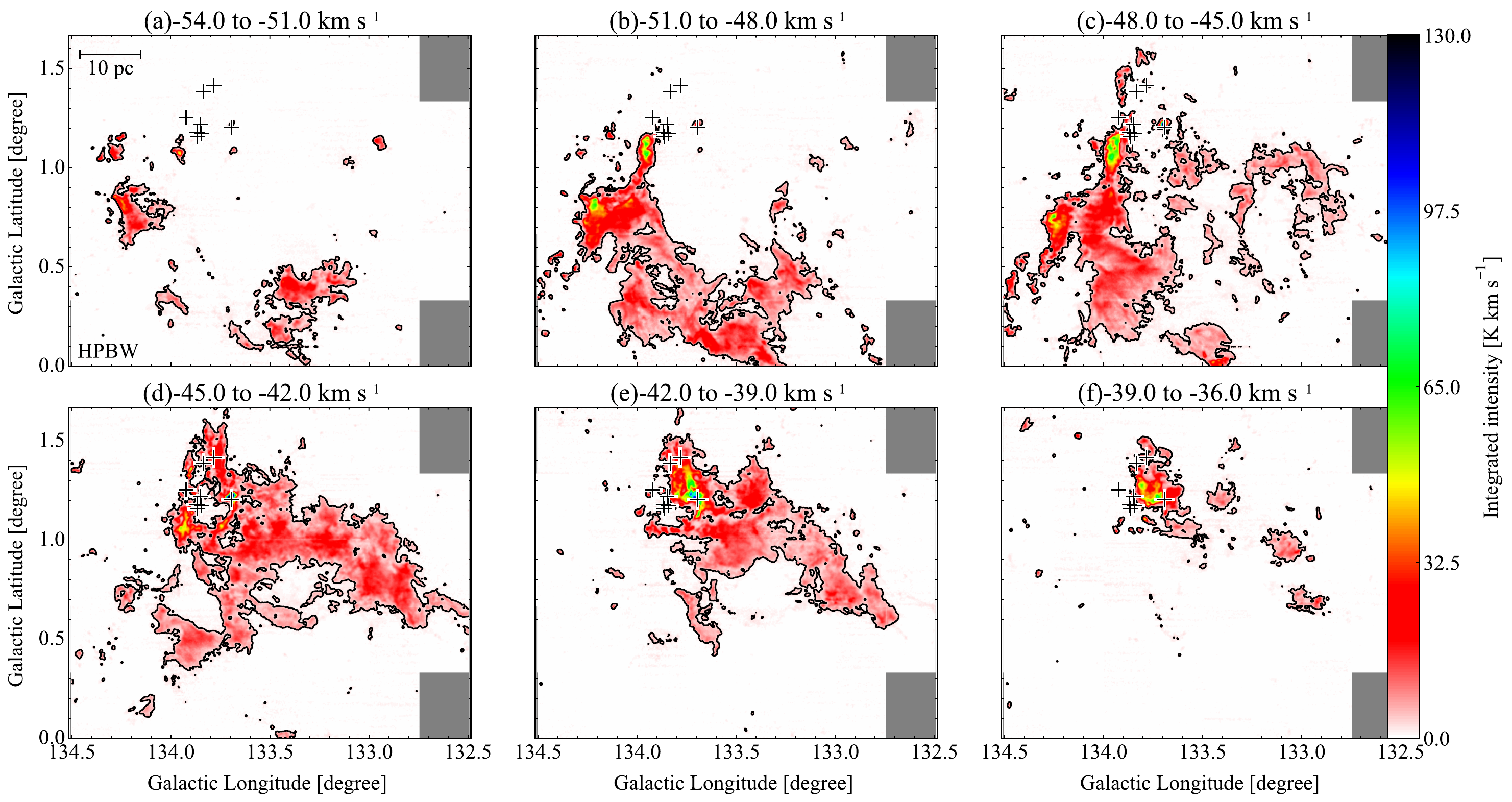}
    \caption{Velocity channel distribution of the $\twelvecoh$ line emission. 
    Integration velocity ranges are denoted at the top of each panel. 
    Superposed contour level is 2.6~$\kkms$. The black crosses represent the positions of O-type stars detected 
    by \citet{Oey_2005}.}
    \label{fig2}
\end{figure*}

Figure \ref{fig1}a depicts a pseudo-colour image taken with \textit{Herschel}\footnote{Herschel is an ESA space observatory with science instruments provided by European-led Principal Investigator consortia and with important participation from NASA.}/PACS at wavelengths of 70~$\mu$m (blue) and 160~$\mu$m (green), along with \textit{Herschel}/SPIRE at 250~$\mu$m (red) \citep{Rivera2013}. The whole W3 GMC is dominated by the red colour at 250~$\mu$m, while the areas around IC~1795, W3 Main, W3(OH), and AFGL~333 are bluer, indicating local dust heating due to nearby high-mass stars. Spectroscopically identified OB stars \citep{Oey_2005} are plotted as black crosses and show correspondence with ``bluer'' regions.

\begin{figure*}
	\includegraphics[width=17cm]{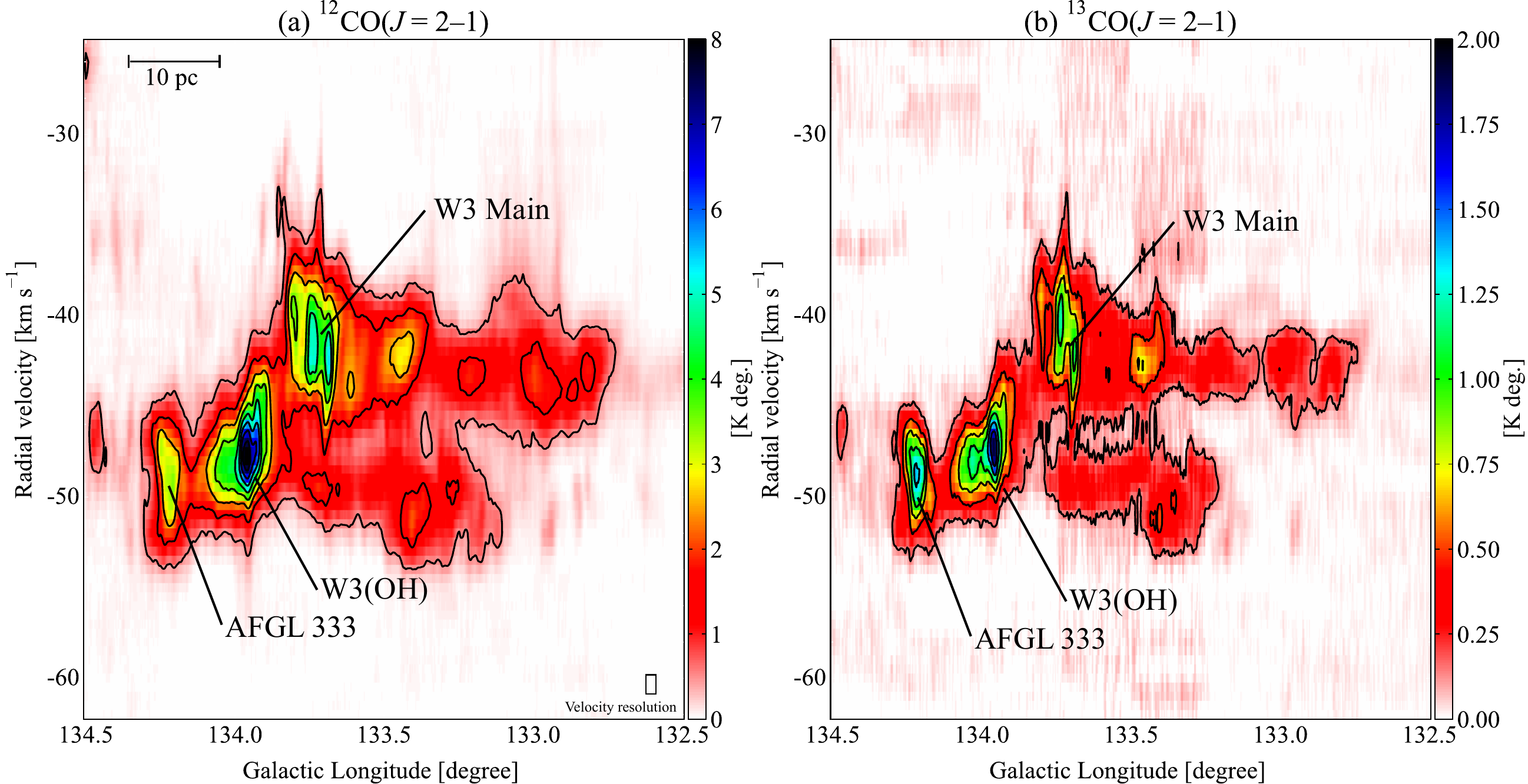}
    \caption{Galactic--longitude velocity diagram at (a) $\twelvecoh$ emission and (b) $\thirteencoh$ emission. The integration ranges for the two panels are $b$=0\fdg0 to 1\fdg67. Contours are plotted every 1.0 K degree from 0.5 K degree ($\sim 16\sigma$) for $\twelvecoh$; every 0.3 K degree from 0.14 K degree ($\sim 7\sigma$) for $\thirteencoh$.}
    \label{fig3}
\end{figure*}

Figure \ref{fig1}b shows the integrated intensity of the $\twelvecoh$ emission. The diffuse gas is extended over the area, and three regions in the northeast-east edge of the cloud show intense $\twelvecoh$ emission, corresponding to the HDL where W3 Main, W3(OH), and AFGL 333 are located \citep{Lada_w3}. W3 Main has the most prominent peak at ($l$, $b$)~$\sim$~(133\fdg70, 1\fdg22) with an integrated intensity of 450~$\kkms$, and most of the O stars in W3 are distributed within 10~pc of the peak position. W3(OH) at ($l$, $b$)~$\sim$~(133\fdg95, 1\fdg07) has an integrated intensity of 300~$\kkms$ and is associated with the OH and H$_2$O masers \citep{Forster,Reid1980} as well as a few OB stars. AFGL~333 has the weakest integrated intensity among the three and is connected to W4 on the eastern edge. The HDL has a clear molecular cavity at ($l$, $b$)~$\sim$~(133\fdg80, 1\fdg13) and coincides with an $\htwo$ region IC~1795 powered by an O6 star called BD+61~411 \citep{Mathys}. The $\htwo$ region corresponds to the ``bluer'' area in Figure \ref{fig1}a, where we can see the intense 70~$\mu$m emission.

Figure \ref{fig2} shows the velocity channel distributions of the W3 region. The distribution shows significant variation in velocity; the gas at $\vlsr$~=~$-54$ to $-45$~$\kms$ (upper panels) is distributed in the southeastern part, while the gas at $\vlsr$~=~$-45$ to $-36$~$\kms$ (lower panels) is distributed in the northwestern region. The two velocity components have different velocity ranges with the boundary at $\vlsr$~$\sim$~$-45$~$\kms$. The three peaks of the $^{12}$CO integrated intensity W3 Main, W3(OH), and AFGL~333 also show different velocity ranges at $\vlsr$~=~$-45$ to $-36$~$\kms$, $\vlsr$~=~$-51$ to $-42$~$\kms$, and $\vlsr$~=~$-54$ to $-45$~$\kms$, respectively.

Figure \ref{fig3} shows Galactic Longitude-velocity diagrams of the W3 region. Figure \ref{fig3}a is for the $\twelvecoh$ emission, and Figure \ref{fig3}b is for the $\thirteencoh$ emission, where the two diagrams show similar distributions. Specifically, W3 Main and W3(OH) show complex velocity distribution. W3 Main shows a velocity difference of 3--4~$\kms$ between $l$ = 133\fdg65 and $l$ = 133\fdg85, and W3(OH) shows a velocity shift of $\sim4$~$\kms$ at $l$ = 133\fdg85 from the peak position at $l$ = 133\fdg95. In the following, we shall focus on the two peaks in W3 Main and W3(OH) and analyse their detailed distributions.

\begin{figure*}
	\includegraphics[width=17cm]{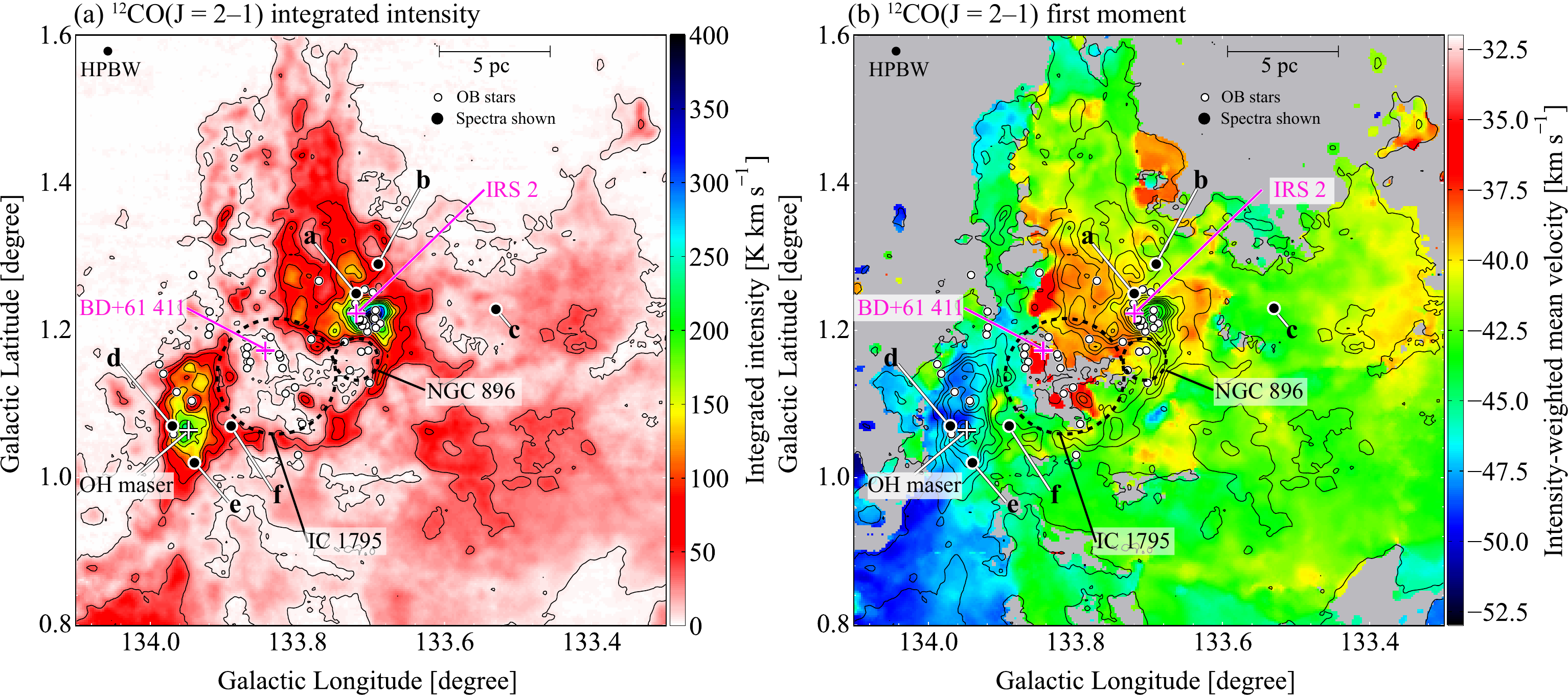}
    \caption{(a) Integrated intensity map of $\twelvecoh$ toward 
    W3~Main. The integration velocity range is from $-55.0$ to $-35.0$~$\kms$.
     The black contours are plotted every 30~$\kkms$ from 5~$\kkms$ for the integrated intensity lower than 216~$\kkms$ and every 60~$\kkms$ from 278~$\kkms$. The cyan 
     dotted circles represent distributions of the $\htwo$ regions IC~1795 and NGC~896. (b) Intensity-weighted mean velocity distributions of the $\twelvecoh$
     emission. The calculation velocity range is from $-55.0$ to $-35.0$~$\kms$. 
     Superposed contours show the integrated intensity distribution, same as a. 
     The black dots a, b, c, d, e and f are the positions where we present spectra in Figure \ref{fig5}. The magenta 
     crosses, white cross, and white dots indicate the positions of BD+61~411, IRS~2, 
     a driving source of the OH maser, and well-known OB stars (e.g., \cite{Navarete2019})
    }
    \label{fig4}
\end{figure*}

\begin{figure*}\centering
	\includegraphics[width=13cm]{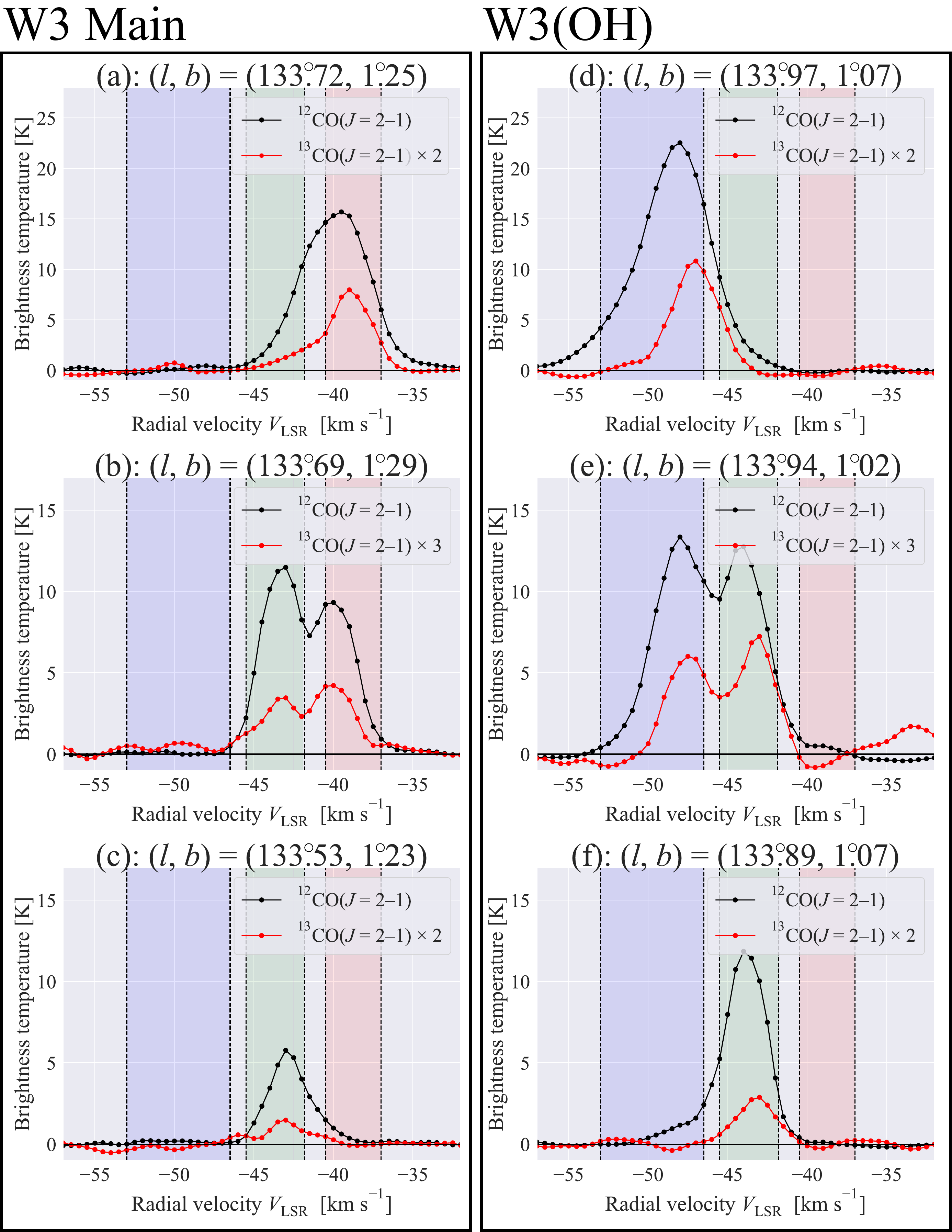}
    \caption{{\it{Left panel}}: $\twelvecoh$ and $\thirteencoh$ spectra in W3 Main region at the positions (a) to (c) in figure \ref{fig4}. {\it{Right panel}}: Same as left panel, but for the W3(OH) region. The blue, green and the red ribbons represent the velocity range of the $-39$~$\kms$ cloud, $-43$~$\kms$ cloud, and $-50$~$\kms$ cloud, respectively.}
    \label{fig5}
\end{figure*}

\begin{figure*}
	\includegraphics[width=17.5cm]{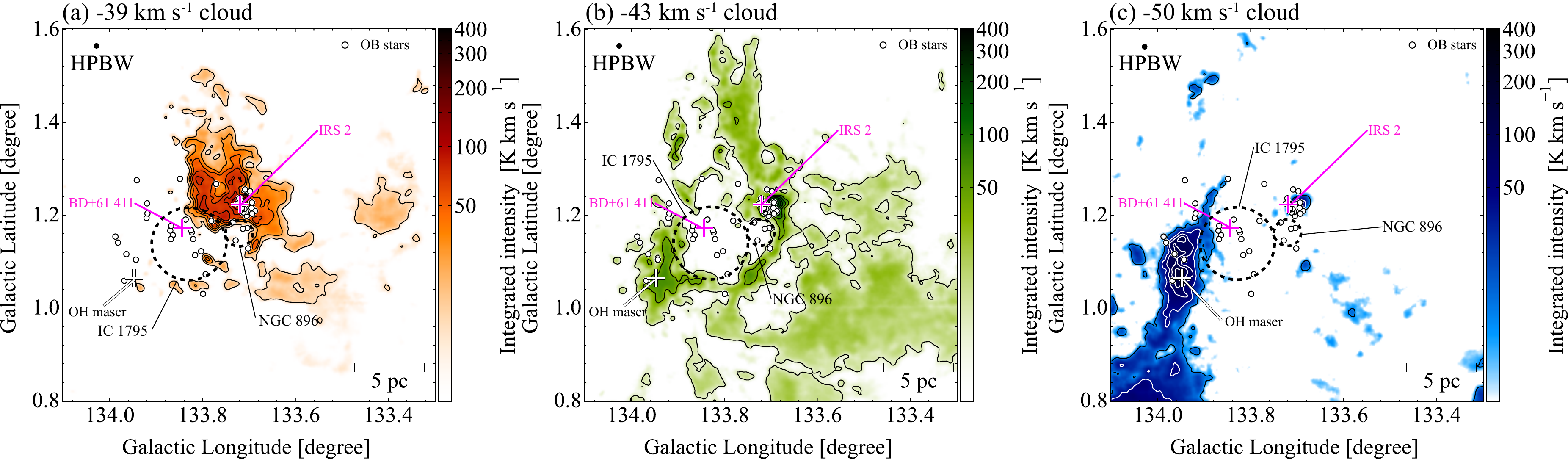}
    \caption{(a) Integrated intensity distribution of the $-39$~$\kms$ cloud. The integration range is from $-40.5$ to $-37.0$~$\kms$. The lowest and the intervals of the superposed contours are 11.7 and 15~$\kkms$, respectively. (b) Integrated intensity distribution of the $-43$~$\kms$ cloud. The integration velocity range is from $-45.5$ to $-42.0$~$\kms$. The lowest and the intervals of the superposed contours are 8 and 30~$\kkms$, respectively. (c) Integrated intensity distribution of the $-50$~$\kms$ cloud. The integration velocity range is from $-53.0$ to $-46.5$~$\kms$. The lowest and the intervals of the superposed contours are 8 and 30~$\kkms$, respectively. The magenta crosses, white cross, and white dots indicate the positions of BD+61 411, IRS~2, a driving source of the OH maser, and well-known OB stars (e.g., \citealp{Navarete2019}).
    }
    \label{fig6}
\end{figure*}

\begin{figure}
	\includegraphics[width=8.5cm]{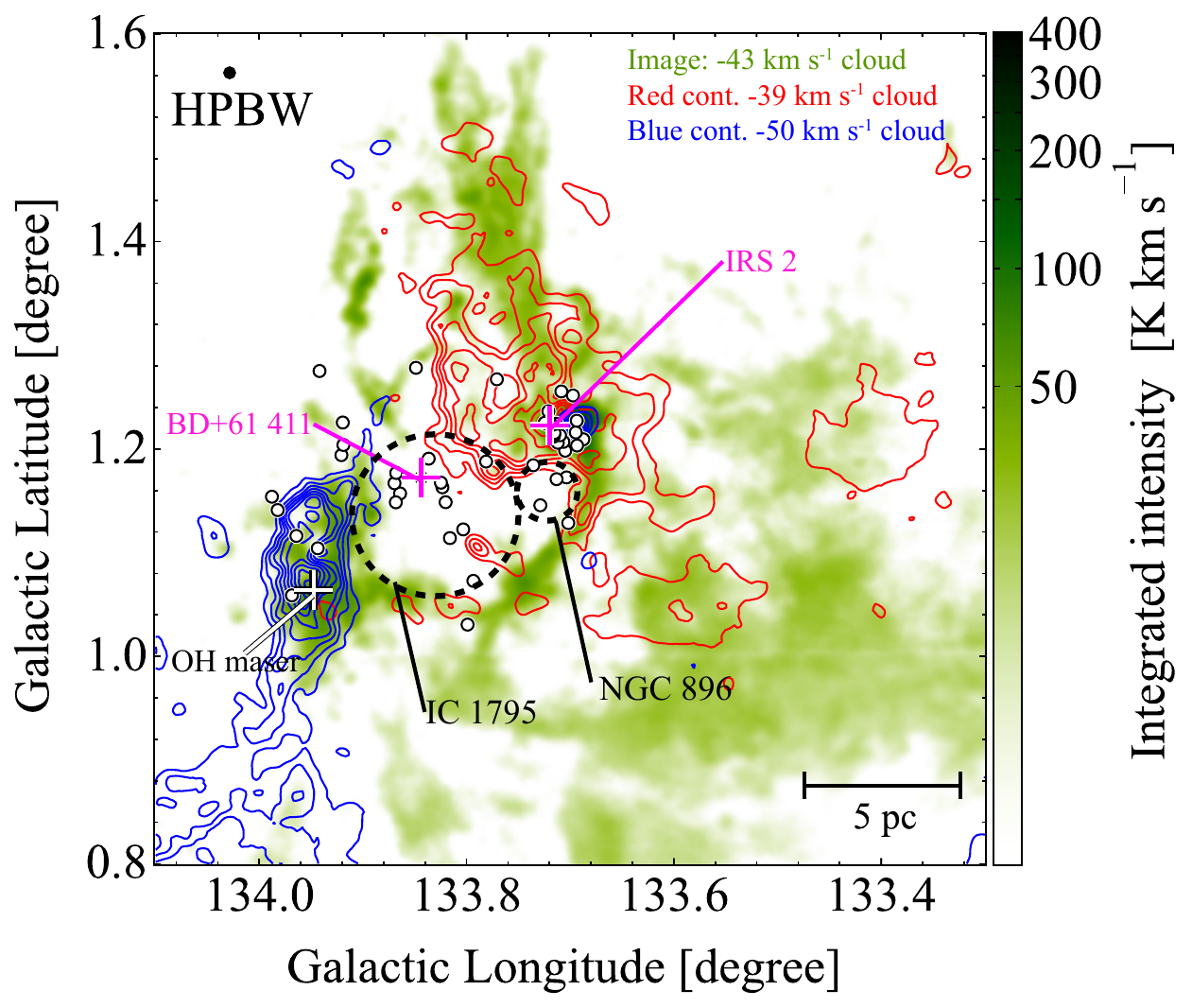}
    \caption{The $\twelvecoh$ distributions of the $-39$~$\kms$ (red contour), $-42$~$\kms$ (image), and $-50$~$\kms$(blue contour) clouds. The contours are plotted every 15~$\kkms$ from 11.7~$\kkms$. The magenta crosses, white cross, and white dots indicate the positions of BD+61 411, IRS 2, a driving source of the OH maser, and well-known OB stars (e.g., \citealp{Navarete2019})}
    \label{fig7}
\end{figure}

Figure \ref{fig4}a shows the integrated intensity distribution of the W3 Main and W3(OH) regions, where the brightest peak corresponds to W3 Main. $\sim$10 O stars (white circles), including BD+61 411 (magenta cross), are concentrated in the W3 Main regions. The $\htwo$ regions IC~1795 and NGC~896 correspond to the cavity of the molecular gas. We see that the OB stars \citep{Navarete2011, Navarete2019, Bik2012} are distributed toward the centre of IC~1795 and W3 Main. BD+61~411(O6.5V), the earliest type member of the IC~1795 cluster, is located inside the molecular cloud cavity toward IC~1795. IRS~2, the most massive star in W3, is associated with W3 Main \citep{Bik2012}. The molecular cavity coincides with these $\htwo$ regions, whereas BD+61 411 has an offset from the centre of the $\htwo$ region overlapping with the cavity. In the W3(OH) region, the molecular emission peak at ($l$, $b$) = (133\fdg95, 1\fdg07) is associated with an OH maser \citep{Forster, Reid1980}. Five O stars are within 5~pc of W3(OH).

\begin{table*}
	\caption{Physical parameters of the molecular clouds}
	\label{tab:1}\centering
	\begin{tabular}{lccr} 
		\hline
		Cloud name& Velocity range & Column density & Mass\\
    & (\kms) & (cm$^{-2}$) & ($\msun$)\\
    \hline
    \hline
        $-39$~$\kms$ cloud & $-40.5$--$-37.0$ & $3.5\times10^{22}$\phantom{0} & $4900$\\
		$-43$~$\kms$ cloud & $-45.5$--$-42.0$ & $6.4\times10^{22}$ & $6800$\\
		$-50$~$\kms$ cloud & $-53.0$--$-46.5$ & $2.7\times10^{22}$ & $3000$\\
		\hline
  \end{tabular}
\end{table*}

Figure \ref{fig4}b shows the distribution of the first moment calculated from a velocity range of $\vlsr$~=~$-55$ to $-35.0$~$\kms$. Only the voxels with intensity greater than 6$\times$$T_\mathrm{rms}$ = 0.72~K are included. Most of the emission is in a velocity range of $\vlsr$~=~$-45.5$ to $-42.0$~$\kms$, whereas red-shifted emission is seen at $\vlsr$~=~$-40.5$ to $-37.0$~$\kms$ with a size of $\sim$5~pc in the east of W3 Main. On the other hand, in the W3(OH) region, the radial velocity of the gas is in a range $\vlsr$~=~$-53.0$~$\kms$ to $-46.5$~$\kms$ on the eastern side, showing a clear velocity shift along the CO cloud elongation of the W3(OH) region.

Figures \ref{fig5}a to 5f show typical CO spectra in the W3 Main and W3(OH) regions, respectively. In the W3 Main region, the CO spectra are peaked in a velocity range of $\vlsr$~=~$-40.5$ to $-37$~$\kms$ and/or $\vlsr$~=~$-45.5$ to $-42.0$~$\kms$. Especially in panel b at ($l$, $b$) = (133\fdg69, 1\fdg29), $^{12}$CO and $^{13}$CO spectra have a double peak in the two velocity ranges, indicating that two clouds are overlapped. Similarly, in the W3(OH) region, the CO cloud has two peaks in velocity ranges of $\vlsr$~=~$-53.0$ to $-46.5$~$\kms$ and $\vlsr$~=~$-45.5$ to $-42.0$~$\kms$, while $^{12}$CO and $^{13}$CO spectra have a double peak in panel e at ($l$, $b$) = (133\fdg94, 1\fdg02), indicating that two clouds are overlapped. The red-shifted component in W3(OH) and the blue-shifted component in W3 Main are part of a common, coherent component, as seen by the first moment (Figure \ref{fig4}). Hence, the W3 Main and W3(OH) regions consist of three clouds, and we hereafter call $\vlsr$~=~$-40.5$ to $-37.0$~$\kms$, $\vlsr$~=~$-45.5$ to $-42.0$~$\kms$, and $\vlsr$~=~$-53.0$ to $-46.5$~$\kms$ components the $-39$~$\kms$, $-43$~$\kms$, and $-50$~$\kms$ clouds, respectively.

Figures \ref{fig6}a, \ref{fig6}b, and \ref{fig6}c show the integrated intensity distributions of the $-39$~$\kms$, $-43$~$\kms$, and $-50$~$\kms$ clouds, respectively. We find that the $-39$~$\kms$ cloud is compact, and its shape fits approximately the cavity of the $-43$~$\kms$ cloud corresponding to IC~1795. The $-50$~$\kms$ cloud is not distributed in the W3 Main region and is concentrated toward the W3(OH) region. Figure \ref{fig7} shows an overlay of the three clouds, and the image of the $-43$~$\kms$ cloud is superimposed over the $-39$~$\kms$ (red contour) and $-50$~$\kms$ (blue contour). The $-39$~$\kms$ and $-43$~$\kms$ clouds overlap in the W3 Main region, while the $-43$~$\kms$ and $-50$~$\kms$ clouds overlap in the W3(OH) region. Thus, the two regions with significantly active star formation are the regions where the two clouds overlap.

\begin{figure}
	\includegraphics[width=9cm]{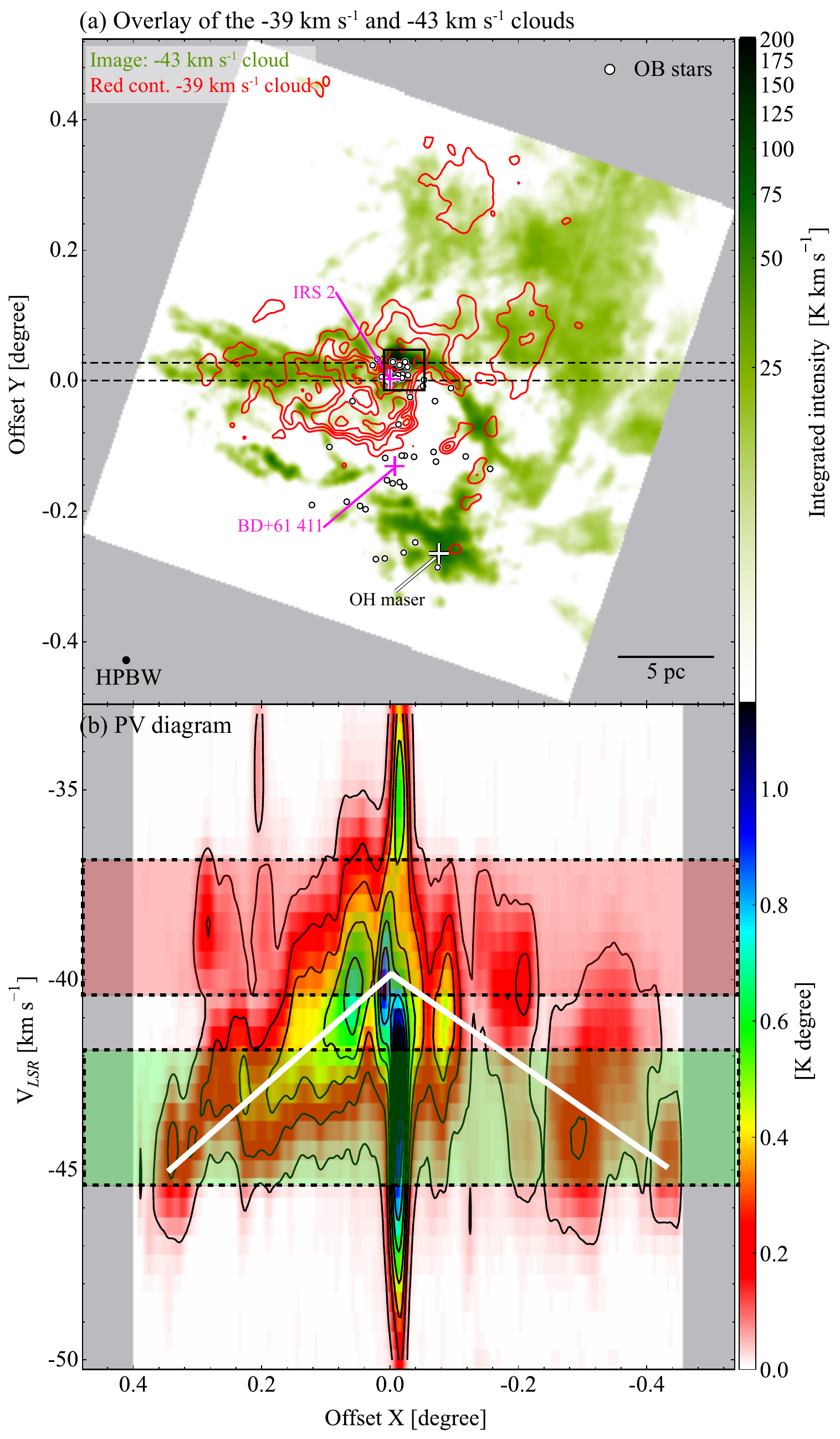}
    \caption{(a) $\twelvecoh$ integrated intensity distribution of the $-39$~$\kms$ and $-43$~$\kms$ clouds in the offset--$X$--$Y$ coordinates. The coordinate is defined by rotating the galactic coordinate counterclockwise by 71\fdg00. The integration velocity range is from $-45.5$ to $-42.0$~$\kms$ for the colour image; $-40.5$ to $-37.0$~$\kms$ for the contours. The lowest and the intervals of superposed contours are 11.7 and 15~$\kkms$. The black box indicates the region to be appeared in (c). The magenta crosses, white cross, and white dots indicate the positions of BD+61 411, IRS~2, a driving source of the OH maser, and well-known OB stars (e.g., \citealp{Navarete2019}) (b) Offset--$Y$-velocity diagram of the W3 Main region in $\twelvecoh$. Integration range along Offset--$Y$ is donated as dotted lines in (a). The velocity ranges of each cloud are represented by red and green transparent belts. The lowest level and the interval of the superposed contours are 0.08 and 0.20~K degrees. The white line indicates a V-shape.}
    \label{fig8}
\end{figure}

Figure \ref{fig8}a shows the overlay of the three clouds, which are rotated by 71 degrees counterclockwise relative to the Galactic plane centred on IRS~2. Figure \ref{fig8}b shows a position-velocity diagram along the horizontal axis in Figure \ref{fig8}a. We find significant line broadening localised toward the CO peak at offset--$X\sim$0.0~degree, indicating blue and red lobes of molecular outflows as mentioned by previous studies (e.g., \citealt{Phillips1988}). The wings extend on both the red-shifted and blue-shifted sides, making it appear bipolar in the present beam, but in reality, the driving sources are different between the Red lobe and the Blue lobe; the Blue lobe is associated with W3 IRS 4, while the Red lobe is associated with IRS 5 \citep{Phillips1988}. In addition to the broad linewidth, differences in the velocities of CO clumps associated with the infrared sources of W3 IRS 4 and W3 IRS 5 have been reported \citep{Thronson1985}. W3 IRS 5 corresponds to the peak at a velocity of approximately $\vlsr$~=~$-40$~$\kms$ for offset--$X\simeq$0.01 degree in Figure \ref{fig8}b, while the peak at offset--$X\simeq$$-0.01$ degree with a velocity of $-43$~$\kms$ corresponds to W3 IRS 4.

\begin{figure}
	\includegraphics[width=8.5cm]{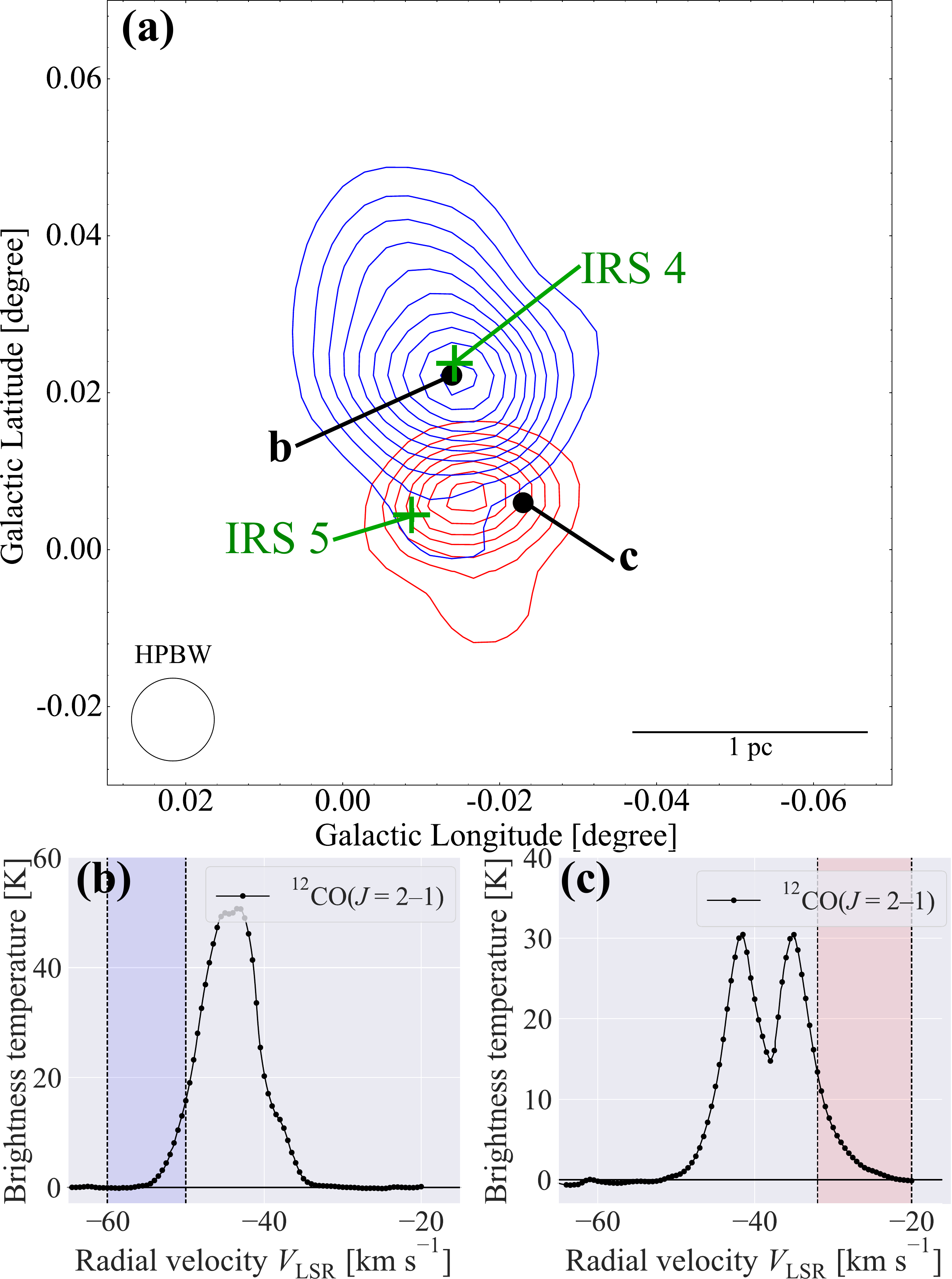}
    \caption{(a) Integrated intensity distribution of the red lobe and the blue lobe are indicated by the red and the blue contours, respectively. The lowest level of the red and the blue contours are 10 and 30~$\kkms$, respectively. The intervals of the red and blue contours are 7 and 20~$\kkms$, respectively. Green crosses indicate the position of IRS~4 and 5. Black dots indicate the position where we show the spectra in panels b and c. (b) Typical spectral profile at position b in (a). (c) Typical spectral profile at position c in (a). The red and the blue ribbons indicate the velocity range of the red- and the blue-shifted wing.}
    \label{fig9}
\end{figure}

Figure \ref{fig9} shows the spatial and spectral distribution of the high-velocity component in the W3 Main region. Regarding W3 IRS 4, observations with a spatial resolution of 3000~au were conducted using IRAM-NOEMA by \citet{Mottram2020}, indicating that it is a hot core associated with a Massive Young Stellar Object (MYSO), and confirming the presence of bipolar outflows. For W3 IRS 5, observations were conducted using the Submillimeter Array (SMA) by \citet{KS-wang}, detecting multiple outflows. Due to insufficient beam resolution, resolving each of these outflows individually is challenging, but line broadening originating from the outflows is localised within approximately 1~pc of W3 IRS 4 and 5. Hence, we find a V-shape by excluding the outflow lobes, as illustrated by the white line in Figure \ref{fig8}b.

\subsection{The physical parameters of the molecular clouds}
In the above sections, we defined three velocity components at $-39$~$\kms$, $-43$~$\kms$, and $-50$~$\kms$ in the HDL, where the most active star formation occurs. We calculated these parameters in the entire region shown in Figure \ref{fig6}.

We derived column densities and masses of each molecular cloud under the local thermodynamic equilibrium assumption. First, we assumed the $\twelvecoh$ line is optically thick and obtained excitation temperature in every pixel. Then, we calculated the optical depth of the $\thirteencoh$ line and derived the $^{13}$CO column density. Next, we obtained molecular hydrogen column densities assuming an abundance ratio of the $^{13}$CO molecules and the H$_2$ molecules of $7.1\times10^5$ \citep{Frerking}. Finally, we obtained masses of the clouds from the column densities. The peak molecular hydrogen column densities toward W3 Main and W3(OH) calculated in the entire velocity cube are $1.12\times10^{23}$~cm$^{-2}$ and $4.1\times10^{22}$~cm$^{-2}$, respectively, giving the total molecular mass of $2.1\times10^4$~$\msun$. The physical parameters of each velocity component are summarised in Table \ref{tab:1}. For details of the analyses, see Appendix. We also confirm that column densities and masses derived in the present study are consistent with those derived from {\it{Herschel}}'s observations within a factor of 1.5 \citep{Rivera2013}. See Appendix for a detailed comparison of physical parameters from the present work and the previous derivation from the {\it{Herschel}}'s observations.

\begin{figure*}
	\includegraphics[width=17cm]{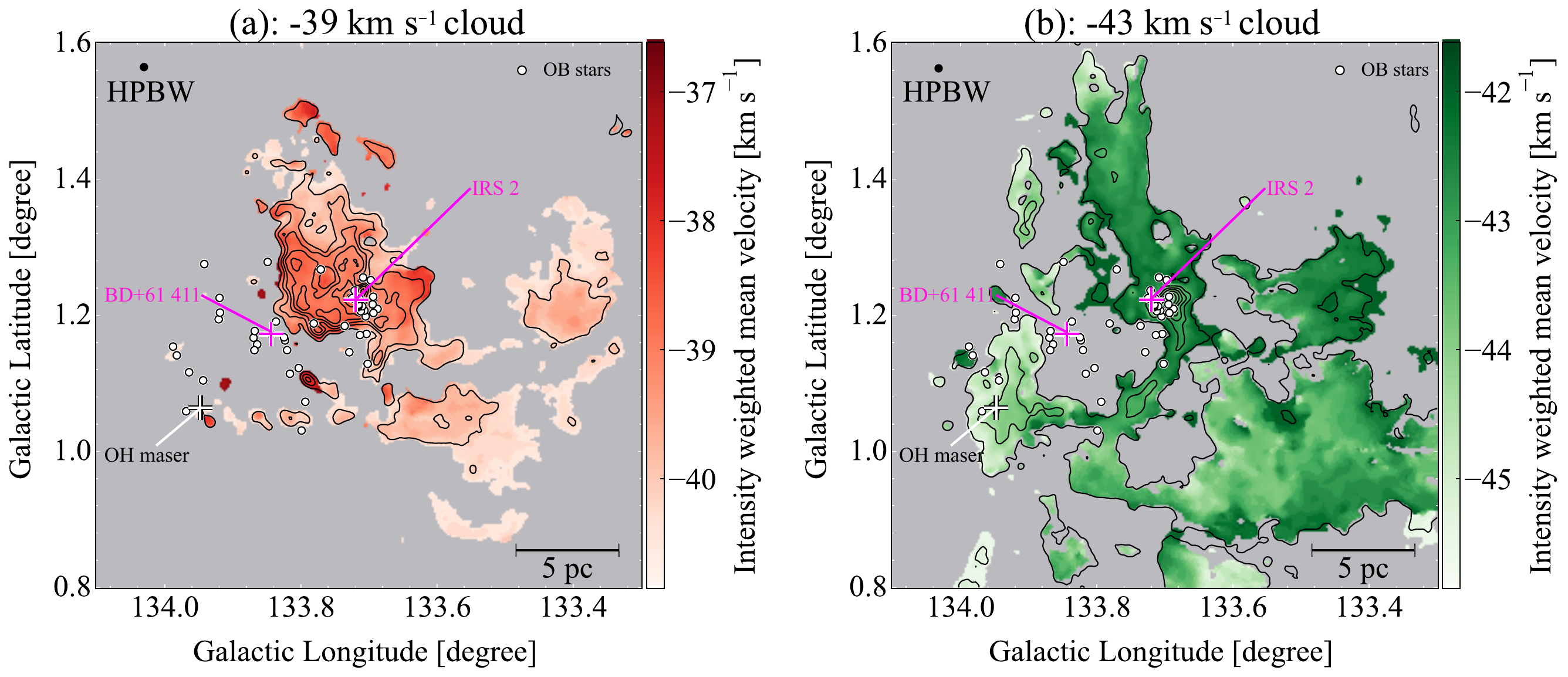}
    \caption{(a) Intensity weighted mean velocity distribution of the $-39$~$\kms$ cloud. 
    (b) Intensity weighted mean velocity distribution of the $-42$~$\kms$ cloud. 
    The contours in the two panels are the same as those of figure \ref{fig6}. 
    The magenta crosses, white cross, and white dots indicate the positions 
    of BD+61~411, IRS~2, a driving source of the OH maser, and well-known OB stars (e.g., \citealt{Navarete2019}).
}
    \label{fig10}
\end{figure*}

%% file: discussion.tex
\section{Discussion}\label{sec:discussion}
Here, we discuss a possible scenario of high-mass star formation in W3 based on a detailed analysis of CO observation.

\begin{figure*}
	\includegraphics[width=17cm]{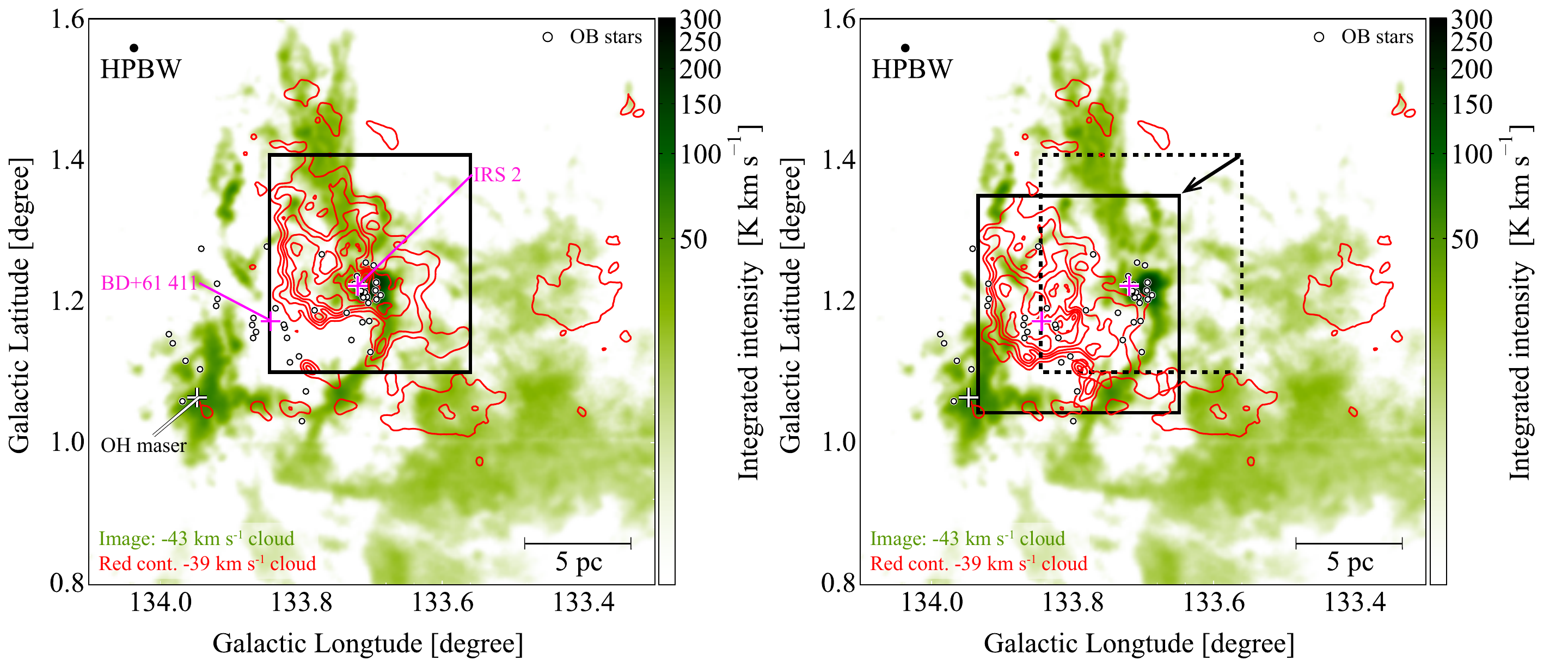}
    \caption{Complementary $\twelvecoh$ distributions of the the $-39$~$\kms$ and $-42$~$\kms$ clouds. The integration velocity range is from 
    $-45.5$ to $-42.0$~$\kms$ for the color image; $-40.5$ to $-37.0$~$\kms$
    for the contours. The lowest and the intervals of superposed contours are 
    11.7 and 15~$\kkms$, respectively. The magenta crosses, 
    white cross, and white dots indicate the positions of BD+61~411, IRS~2, 
    a driving source of the OH maser, and well-known OB stars (e.g., \citealt{Navarete2019}).
}
    \label{fig11}
\end{figure*}

\subsection{Trigger of star formation by the expansion of the $\htwo$ region}
The scenario of sequential star formation driven by OB associations was proposed and 
discussed by \cite{Elmegreen_Lada}. In the W3 region, \cite{Lada_w3} presented 
a model that the W4~$\htwo$ region expanded to compress the gas to form the HDL in the W3 
region which led to trigger star formation. Subsequently, \cite{Oey_2005} proposed a 
``Hierarchical triggering'' model where the expansion of the Perseus chimney/super bubble 
in W4 triggered the formation of the IC~1795 cluster, and the feedback of the IC~1795 cluster triggered the active star formation in W3~Main and W3(OH). These previous studies are based on the stellar-age distribution, whereas detailed kinematics and distribution of the parental molecular gas were not analysed/considered in depth. We aim to better understand the star formation by using the velocity and density distribution of the CO molecular gas.

We find two molecular clouds in the W3~Main region (Section \ref{sec:results}). The two clouds have 
a velocity difference of $\sim$4~$\kms$, and a usual explanation for the velocity difference 
is gas acceleration by the feedback of high-mass stars. In the present case, a possible scenario 
is that the momentum released by the IC~1795 cluster is responsible for the acceleration of the $-39$~$\kms$ cloud relative to the $-43$~$\kms$ cloud. Because the $-39$~$\kms$ cloud seems to be accelerated by $\sim$4~$\kms$ from the systemic velocity if we regard the V-shape structure in the position velocity diagram as a red-shifted half of the expanding shell (see figure \ref{fig8}b). Further, the 
$-39$~$\kms$ cloud has an elongated structure toward BD+61~411, which may correspond to a blown-off cloud efficiently overtaken by the stellar wind (e.g., \citealp{Fukui_elongated_structure, Sano_elongated_structure}). To evaluate the expansion scenario, we first consider the stellar winds as the momentum source. The highest-mass star in 
IC~1795 is an O6.5V star BD+61~411 \citep{Oey_2005}. The momentum delivered 
by the star to the surroundings in a time span of 3--5~Myr, the age of 
the cluster \citep{Oey_2005}, is expressed as follows;
\begin{equation}
P_{\mathrm{winds}}=\dot{M}v\cdot\Delta t
\end{equation}
where $P_{\mathrm{winds}}$, $\dot{M}$, and $\Delta t$ are the wind momentum, the mass loss rate, and the cluster age, respectively. The typical wind mass loss rate of an O6.5V star is $10^{-6.4}$~$\msun$ yr$^{-1}$ \citep{Vink} and its terminal velocity is 2600~$\kms$ \citep{Kudritzki}. If we assume that 
IC~1795 contains ten O6.5V stars, we estimate $P_\mathrm{winds} = 3.1 \times 10^4$~$\msun$~$\kms$. 
The momentum required to accelerate a cloud of $M_\mathrm{cloud}$ to a relative velocity 
$V_\mathrm{rel}$ is given as follows if a complete coupling of the momentum to the cloud mass 
is assumed; 
\begin{equation}
P_{\mathrm{clouds}}=\dot{M}_\mathrm{clouds}V_\mathrm{rel}
\end{equation}

By taking the cloud mass accelerated by 4~$\kms$ to be $1.17\times10^4$~$\msun$ ($-39$~$\kms$ cloud \& $-43$~$\kms$ cloud), 
we obtain $P_\mathrm{cloud} = 4.68 \times 10^4$~$\msun$~$\kms$. Similarly, in the W3(OH) region, we obtain $P_\mathrm{cloud} = 6.86 \times 10^4$~$\msun$~$\kms$ assuming the gas with a mass of 9800~$\msun$ ($-43$~$\kms$ cloud \& $-50$~$\kms$ cloud) being accelerated by 7~$\kms$. These values are 
comparable to that supplied by the stellar winds. However, 
we need to further consider the realistic geometry of the clouds and 
the stars. The solid angle subtended by the cloud relative to 
IC~1795 seem to be significantly less than 4$\pi$ due to the 
inhomogeneous distribution of the clouds as shown by the weak CO emission toward the IC~1795 cluster (Figure \ref{fig6}a). This indicates that the stellar feedback is hardly able to explain the cause of the velocity difference.

It has been a usual thought for a long time that the $\htwo$ regions can accelerate and 
compress the surrounding gas. This possibility was studied theoretically by \cite{Kahn1954}
and it was shown that there is a solution corresponding to the acceleration phase of 
the neutral gas driven by the $\htwo$ gas under a range of two parameters: the stellar 
ultraviolet photons and the gas density. By hydrodynamical numerical simulations 
without magnetic field, \cite{Hosokawa_Inutsuka} showed that the gas around an 
O star is accelerated and compressed, leading to star formation. Subsequently, 
\cite{Krumholz_2007} made three-dimensional magnetohydrodynamical numerical simulations of the 
interaction between an $\htwo$ region and magnetised neutral gas. They showed that 
the $\htwo$ region expands in a direction parallel to the magnetic field lines, 
whereas the expansion is strongly suppressed in the direction perpendicular to the 
field lines by the magnetic pressure. They concluded thus that the triggered star 
formation is much less effective than thought previously without a magnetic field. 
It is also known that the Champagne flow makes the $\htwo$ gas escape from a low-density 
part of the surrounding gas (\citealp{Tenorio-Tagle}; see Figure 4 of \citealp{2008hsf1.book..264M}), 
leading to even less compression. W3 seems to be one of such a case. In summary, 
recent numerical studies of the interaction of a $\htwo$ region with the ambient 
neutral gas show that a $\htwo$ region may not play a role in gas compression.

We are, in fact, able to see a trend of no acceleration of molecular gas in the present 
data. Figure \ref{fig10} shows the first moment distributions of the $-39$~$\kms$ and $-43$~$\kms$
clouds. The $-43$~$\kms$ cloud in Figure \ref{fig10}b shows that there is no appreciable velocity shift of the molecular gas facing the $\htwo$ region at the present resolution, which contradicts 
the previous numerical simulations without magnetic field predicting a velocity 
shift (see Figure 1 of \citealp{Hosokawa_Inutsuka}). The $-43$~$\kms$ cloud has shell-like 
distribution and has the first moment at $-43$~$\kms$ over the whole extent with insignificant velocity 
variation. This may suggest that, while the cloud has a shell morphology, feedback by 
the IC~1795 cluster has a small effect on the overall cloud kinematics/distribution. 
On the other hand, the $-39$~$\kms$ cloud shows a small velocity variation less than $\sim$1.5~$\kms$ in some 
places.
Part of it may be due to the feedback of the stars, while the effect does 
not seem to be prevailing. 

In summary, it seems that the usual scenario of feedback by high-mass stars is 
not able to explain the present observations of the W3 molecular gas in triggering 
star formation. This is along the line which has been argued for in recent works, 
whereas the trigger process by stellar feedback in star formation has not yet been fully 
tested including the effects of small-scale clumps. We note that small-scale triggering star formation at ``sub-pc scale'' has been discussed in the previous works \citep{Tieftrunk1997,KS-wang,Rivera2013}, while we do not deal with them further in the present work of pc-scale resolution.

\subsection{An alternative scenario; Cloud-cloud collisions}\label{subsec:ccc}

\begin{figure}
	\includegraphics[width=9cm]{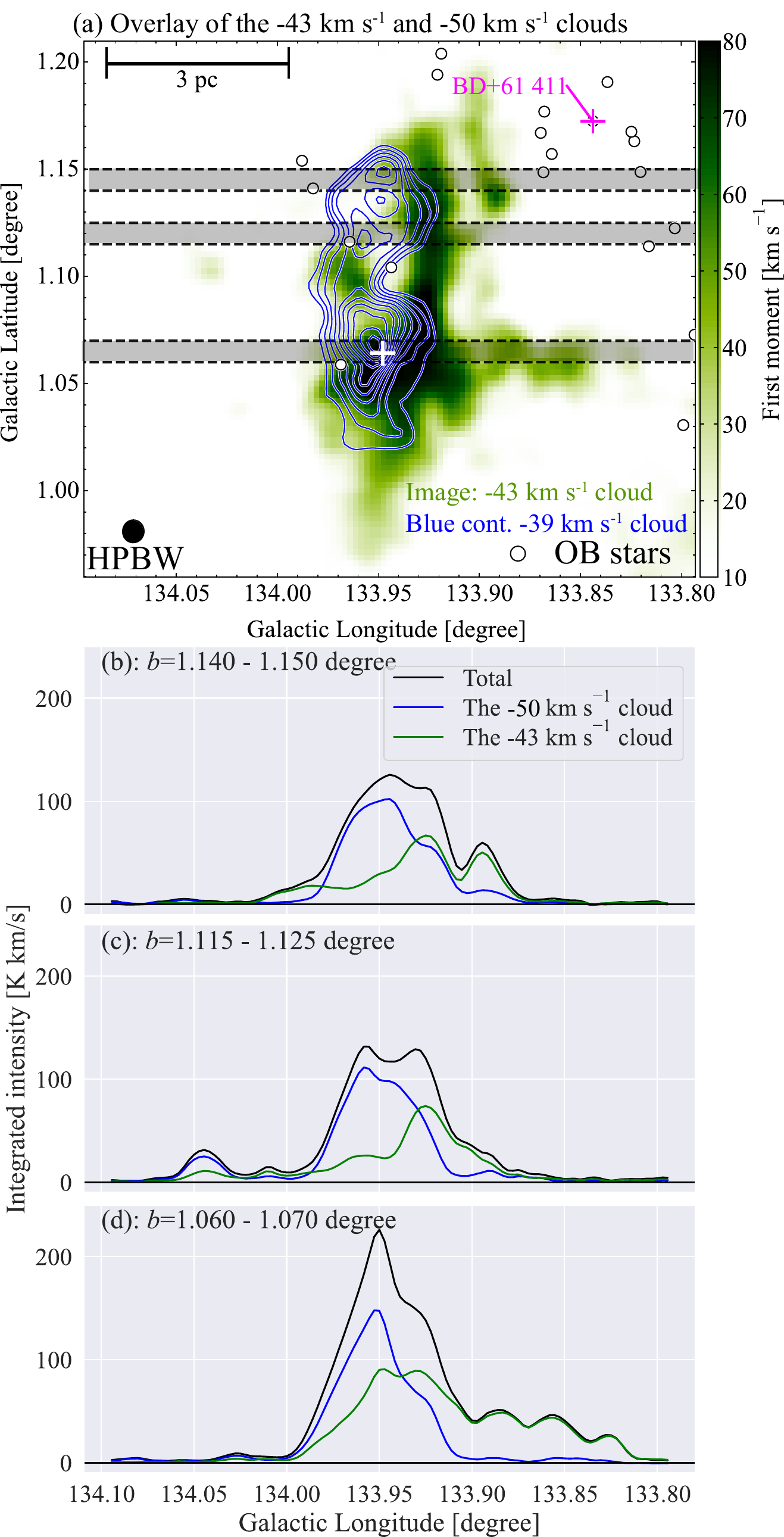}
    \caption{(a) $\twelvecoh$ distributions of the $-50$~$\kms$ and $-43$~$\kms$ clouds. The integration velocity range is from $-53.0$ to $-46.5$~$\kms$ for the colour image; $-46.5$ to $-42.0$~$\kms$ for the contours. The lowest and the intervals of the superposed contours are 60 and 10~$\kkms$. (b)--(d) Integrated intensities of each cloud averaged along the galactic latitude. Positions of each profile are indicated as dark transparent belts enclosed by the dotted lines in (a). The magenta crosses, white cross, and white dots indicate the positions of BD+61 411, a driving source of the OH maser, and well-known OB stars (e.g., \citealt{Navarete2019})}
    \label{fig13}
\end{figure}

\begin{figure}
	\includegraphics[width=8cm]{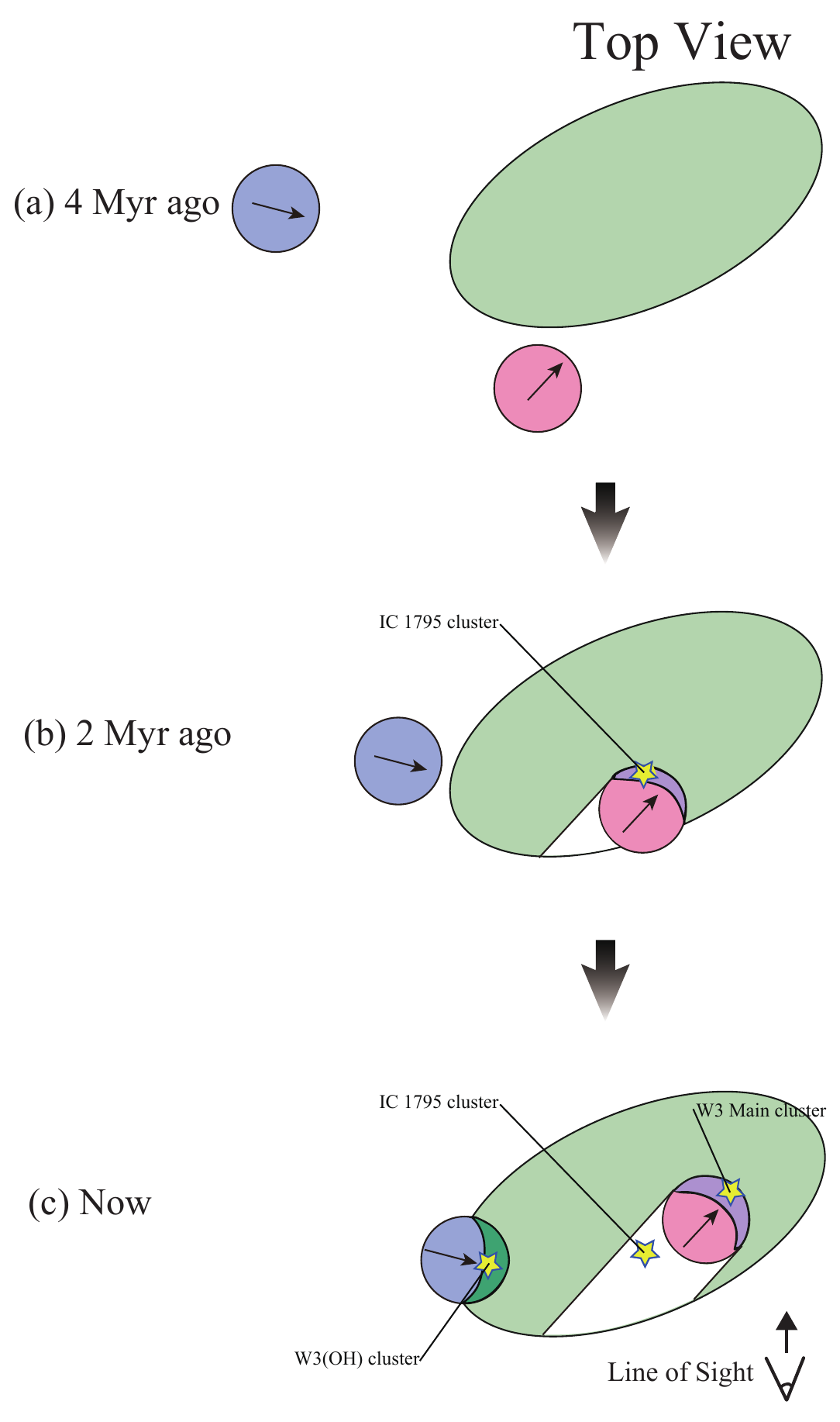}
    \caption{Schematic images of the cloud-cloud collisions in the W3 for the phases of (a) before the collision, (b) the collision in progress, and (c) the present day. The red, green, and blue colours represent the $-39$~$\kms$, $-43$~$\kms$, and $-50$~$\kms$ clouds, respectively. Arrows on the red and blue colours show the direction of movement of the $-39$~$\kms$ and $-50$~$\kms$ clouds, respectively. }
    \label{fig14}
\end{figure}

We discussed it is unlikely that the sequential star formation model is able to 
explain the star formation in W3 Main and W3(OH). In this section, we test a 
scenario based on CCCs, which are suggested to be 
operating in the other typical young $\htwo$ regions including the Orion region, 
the Carina region, and the M16 and M17 region \citep{Fukui_rev,Fukui_2018_Orion,Enokiya_2021,Fujita_carina,Nishimura_2018,Nishimura_2021}.

\cite{Fukui_rev} summarized the main observational signatures of a CCC, 
where two clouds with a supersonic velocity separation associated with a young 
star/cluster or their candidates as follows;

\begin{enumerate}[i]
    \item complementary spatial distribution between the two clouds, and
    \item bridge feature(s) connecting the two clouds in velocity space, and
    \item U-shaped morphology.
\end{enumerate}
These three features are based on numerical modelling of colliding clouds by \citet{Habe_Ohta} and subsequent observational tests summarised by \citet{Fukui_rev}. This model assumes the collision between two clouds of different sizes, and the smaller cloud penetrates deeply into the large cloud and creates a cavity in the large cloud. Because this cavity is made by the motion of the small cloud, the small cloud and cavity show spatially complementary (or anti-correlated) distribution. In a case where the collision direction and line-of-sight have a non-zero angle, the complementary distribution between the small cloud and cavity has displacement due to the projection effect. This displacement is predicted by the numerical calculation of \citet{Takahira2014} and its synthetic observation by \citet{Fukui_2018_Orion}, and actually observed (e.g., \citealp{Fukui_2018_Orion, Sano_2021_M33, Fujita_carina, Yamada2021}). 

The colliding two clouds exchange momentum, and an intermediate velocity component arises, which is observed as the bridge feature in the position-velocity diagram. The bridge feature is often observed as a ``V-shape'' in the position velocity diagram with the small cloud at the vertex and two bridges linking to the large cloud. The U-shape is due to the compression of the large cloud induced by the penetration of the small cloud. Because of the compression, the bottom of the ``U-shape'' is the densest in the colliding two-cloud system.

In W3 Main, we argue that the properties of the $-39$~$\kms$ cloud and the $-43$~$\kms$ cloud show the three observed signatures i--iii of CCC. 4~$\kms$, a projected velocity separation, gives a lower limit to the actual collision velocity and is larger than the typical linewidth of the two clouds. A realistic local sound speed is provided by the Alfvenic speed, which is a few $\kms$ if density and magnetic field are assumed to be 300 cm$^{-3}$ and 20 $\mu$G, respectively, being the typical conditions of the interstellar medium \citep{Crutcher2010}. The two clouds peaked toward IRS 2 in the cluster identified by \citet{Bik2012}, which corresponds to the densest part in the U-shaped morphology and matches the picture of a U-shape formed by the Habe-Ohata model. 

Figure \ref{fig11}a shows an overlay of the $-39$~$\kms$ and the $-43$~$\kms$ clouds. The $-43$~$\kms$ cloud is extended beyond the outside of W3 Main as diffuse emission, while the $-39$~$\kms$ cloud has a clear boundary of a 5 pc diameter. In order to test if a cloud-cloud collision is a possible scenario, we applied the algorithm of displacement by \citet{Fujita_carina}. This code optimises the correlation coefficient between the $-39$~$\kms$ cloud and the $-43$~$\kms$ cloud enclosed by the two boxes in Figure \ref{fig11}a. We then obtained the lowest correlation coefficient of $-0.53$ at the displacement of 4.1~pc to the southeast, achieving a good matching with the cavity in the large cloud. The correlation coefficient of $-0.53$ is appropriate to fit the displacement (e.g., \citealp{Fujita_carina, Sano_2021_M33}). Figure \ref{fig8} shows a V-shape in the position velocity diagram as indicated by the white line superposed. The $-39$~$\kms$ cloud is more compact than the $-43$~$\kms$ cloud; the top of the V-shape is extended more toward the $-39$~$\kms$ cloud. 

We suggest that the W3(OH) region is also explicable by a CCC scenario. In this region, the $-50$~$\kms$ and $-43$~$\kms$ clouds overlap and are associated with an OH maser source and an H$_2$O maser source as well as a few B stars. Figure \ref{fig13}a shows a complementary distribution between the two 
clouds. Figures \ref{fig13}b--\ref{fig13}d show a strip map of Figure \ref{fig13}a, which presents the 
integrated intensity in the dark shaded area averaged in Galactic Latitude. The blue, green, and black lines show the $-50$~$\kms$, $-43$~$\kms$, and whole W3(OH) cloud, respectively. We find that the $-50$~$\kms$ cloud and the $-43$~$\kms$ cloud are overlapping with each other at $l$ = 133\fdg90--133\fdg95, and suggest it is possible that the two clouds are merging via collision there. In section \ref{CCC_W3OH}, we discuss the detailed scenario of CCC in W3(OH).

\subsection{A possible CCC scenario}
Based on the CCC signatures in Section \ref{subsec:ccc}, we assume that CCCs are operating in the W3 Main and W3(OH) regions and discuss the high-mass star formation of IC~1795, W3 Main, and W3(OH).

Figure \ref{fig14} summarises a star formation scenario based on cloud-cloud collisions. The $-43$~$\kms$ cloud, the most massive among the three, is associated with three clusters W3 Main, IC~1795, and W3(OH), while the gas around IC~1795 has already dispersed due to stellar feedback. The two clusters, IC~1795 and W3 Main, are located along the path of the $-39$~$\kms$ cloud estimated by the CO distribution and radial velocity, which are independent of the cluster information. According to the optical and infrared observations, the IC~1795 cluster is older than that of the W3 Main cluster. This is consistent with a scenario that the $-39$~$\kms$ cloud collided with $-43$~$\kms$ at two different epochs, where the first collision triggered IC~1795 and the second collision triggered W3 Main. The other cluster, W3(OH), is located at the eastern edge of the $-43$~$\kms$ cloud, separated by more than 10~pc from the two clusters above and not along the path of the $-39$~$\kms$ cloud. Hence, it is hard to consider that the $-39$~$\kms$ cloud affects the star formation in W3(OH). On the other hand, and the eastern part of the $-43$~$\kms$ cloud shows signatures of a cloud-cloud collision with the $-50$~$\kms$ cloud. Hence, we suggest that the W3(OH) cluster was formed by the collision between the $-50$~$\kms$ and $-43$~$\kms$ clouds recently.

\subsubsection{Multiple episodes of star formation triggered by the $-43$~$\kms$ and $-39$~$\kms$ clouds}
IC~1795 and W3 Main have different ages as shown by the previous works. IC~1795 is associated with an $\htwo$ region, and most of its members are visible at optical wavelength. The age of the cluster was studied by the HR diagram \citep{Ogura_Ishida, Oey_2005, Kiminki2015}, and the stellar ages are usually suggested to be 3--5 Myr. However, the age determination based on the HR diagram has uncertainties, mainly because of the assumption of extinction law and stellar evolution model. According to the $J$, $H$, and $K$ band observations by Calar Alto Observatory 3.5-m observation, IC~1795 still hosts a number of young sources having circumstellar disks, indicating the star formation was active at least in the recent 2 Myr \citep{Roman_Zuniga}. These optical and infrared studies suggest that the star formation in IC~1795 continued until 2 Myr ago, indicating a cluster age younger than 3 Myr. On the other hand, W3 Main includes very young objects as evidenced by the molecular outflow sources having a dynamical age of $10^4$~yr, and the formation of W3 Main is very recent, within $10^5$~yr. The two clusters were, therefore, formed in two different epochs separated by 2 Myr.

Based on the age separation above, we present a scenario in which the two clusters were formed by two trigger events separated by 2 Myr. Figure \ref{fig14} shows a schematic illustration of the possible cloud motion and three-dimensional cloud structure in the past 4 Myr. The $-39$~$\kms$ clouds collided with the $-43$~$\kms$ cloud 2 Myr ago and continued to penetrate into the $-43$~$\kms$ cloud until now. Because we estimated the projected displacement of 4.1 pc by the velocity difference of $4$~$\kms$ (Figure \ref{fig11}), the collision time scale is given as follows, 

\begin{equation}
    t = (4.1 \mathrm{[pc]}/4 \mathrm{[km/s]})\cdot(1/\mathrm{tan}\theta)\sim1\cdot(1/\mathrm{tan} \theta)   [\mathrm{Myr}],
\end{equation}

where $\theta$ is the angle between the sightline and the cloud velocity vector. If we assume $\theta$ to be less than $30$ deg, the collision time scale becomes $\sim$2 Myr as is consistent with the age of the IC~1795 cluster. Although the exact mass of the IC~1795 cluster is not precisely determined, the most massive star has a spectral type of O6.5 V((f)) whose mass is calculated to be around 30~$\msun$ according to the calibration by \citet{Martins2005}. The mass distribution of the IC~1795 cluster is consistent with Kroupa’s Initial Mass Function (IMF) as shown by \citet{Roman_Zuniga}. By assuming the IMF, the cluster mass is estimated to be around 1000~$\msun$. The Star Formation Efficiency (SFE) is derived as SFE = (Cluster Mass) / (Cluster Mass + Molecular Cloud Mass) = 1000 / (1000+4900) = 17\%, where molecular mass corresponds to the mass of the $-39$~$\kms$ cloud in the present scenario. The SFE is likely an upper limit because the mass of the $-43$~$\kms$ cloud at the position of IC~1795 was likely dissipated due to the formation of IC~1795 and the ionisation, implying that the $-43$~$\kms$ cloud was more massive prior to the collision than it is at present. 

We suggest that the collision is still ongoing and is triggering high-mass star formation in W3 Main. The $-39$~$\kms$ and $-43$~$\kms$ clouds overlap toward W3 Main (Figure \ref{fig7}), corresponding to phase 3 in Figure 1 of \citet{Fukui_rev} and the compressed gas has a total column density of over $10^{23}$~cm$^{-2}$. The gas distribution indicates that W3 Main corresponds to the shock-compressed merged gas in the Habe \& Ohta model. This idea is supported by the presence of molecular outflows indicating star formation with a timescale of $10^4$~yr. In addition to outflows, various stages of high-mass star formation, including the diffuse $\htwo$ region W3 H, J, K, compact $\htwo$ region W3 A, B, and D, Ultra-compact $\htwo$ region W3 C, E, F, and G, as well as Hyper compact $\htwo$ region W3~Mag and Ca, which were identified previously \citep{Tieftrunk1997}, are embedded in around 2~pc from the CO peak toward W3 Main. Moreover, \citet{Troland1989} measured the magnetic field strength of W3 Main to be $\sim$100~$\mu$G using the Zeeman splitting method based on the $\hone$ observation by VLA. Such a high magnetic field strength is consistent with MHD simulations of \citet{Inoue_Fukui} in which the collision can compress the gas to enhance the magnetic field up to $\sim$100--600~$\mu$G, leading to larger Alfvenic velocity and effective Jeans mass. Because the total stellar mass of the W3 Main region was estimated to be 4000~$\msun$, the SFE is calculated to be 25\% by taking the molecular cloud mass to be $1.17\times10^4$~$\msun$ ($-43$~$\kms$ cloud \& $-39$~$\kms$ cloud). Another small cluster, NGC~896, is located in the south of W3 Main. The age of the cluster is estimated to be 1--2 Myr by \citet{Roman_Zuniga} and is likely formed between IC~1795 and W3 Main. We infer that the formation of NGC~896 was triggered by a collision between the $-43$~$\kms$ cloud and the western part of the $-39$~$\kms$ cloud $\sim$1 Myr ago, nearly continuously following the first collision which triggered IC~1795 (see Figure \ref{fig14}).

\subsubsection{The collision between the $-50$~$\kms$ and $-43$~$\kms$ clouds in the W3(OH) region}\label{CCC_W3OH}

Star formation in W3(OH) seems less active than that in W3 Main. We next argue that the high mass stars W3(OH) were formed 0.7 Myr ago by the collision between the $-43$~$\kms$ cloud and the $-50$~$\kms$ cloud. The molecular clouds in the W3(OH) region are elongated along the Galactic plane. Figure \ref{fig13} shows that the $-43$~$\kms$ and $-50$~$\kms$ clouds have a similar size, which suggests that the two clouds collided with each other to form the elongated cloud as observed. Instead of the picture by \citet{Habe_Ohta} where the small cloud compresses the large cloud to create a cavity in it, as in Figure \ref{fig13}, the two clouds with similar extents perpendicular to the collision direction collided to compress gas into an elongated structure along the Galactic latitude. Such a mode of CCC was reported in the Sh2-233 region (See Figure 6 of \citealp{Yamada2022}).

We estimate the time scale by the crossing time from 3 pc/4~$\kms$ = 0.7 Myr because there is no hint of a displacement between the two clouds. Numerical simulations of CCCs \citep{Habe_Ohta, Anathpindika, Takahira2014} show that star formation does not take place just after the collision onset but some time after the merging begins. Therefore, the real star formation happened well after the time scale. The ages of the maser sources, W3(OH) and W3(H$_2$O), are roughly consistent with the timescale of the collision. The driving sources of OH maser are stars of the Zero Age Main Sequence (ZAMS) whose age is $\sim0.1$ Myr. Further, the driving sources of H$_2$O masses whose age is a few 10$^3$ yr are hot cores \citep{Sheng-Li_2016}. The W3(OH) region star formation is ongoing with merging clouds. The short lifetimes do not contradict the CCC model. 

Considering the above, we frame a scenario as follows; in W3(OH), the $-50$~$\kms$ and the $-43$~$\kms$ clouds began to collide about 0.7 Myr ago along the Galactic plane. The two clouds have a column density of $\sim10^{22}$ cm$^{-2}$. This filamentary molecular cloud became gravitationally unstable in the last 10$^5$ yr and triggered the formation of the OH maser and H$_2$O maser. The W3 (OH) region has a number of B stars whose age is 2--3 Myr \citep{Roman_Zuniga}. They are unrelated to the collision of $-43$~$\kms$ and $-50$~$\kms$ clouds.

Up to this point, we have discussed the potential scenarios of CCC in both the W3 Main region and the W3(OH) region. As illustrated in Figure \ref{fig14}, both regions are associated with the same $-43$~$\kms$ cloud. In the case of the W3 Main region, star formation could be induced by the collision with the $-39$~$\kms$ cloud, while in the W3(OH) region, it could be triggered by the collision with the $-50$~$\kms$ cloud. Furthermore, the formation of IC 1795 can also be explained depending on the assumed collision angles. 

The W3 GMC has a molecular mass of 10$^6$~$\msun$ along with the associated stellar mass of over 5000~$\msun$. The molecular mass is outstanding in the outer solar circle. The GMC consists of three velocity components. The mass of the $-43$~$\kms$ cloud, the most massive among the three, is 6800~$\msun$, and its collision with the $-39$~$\kms$ cloud is a usual CCC according to the simulations of the collisions in the Galactic plane (\citealt{Kobayashi2017}; see also Figure 9d of \citealt{Fukui_rev}).

\subsection{Comparison with the other CCC candidates}
It is known that W3~Main has a Trapezium-like cluster \citep{Abt2000}. 
Table \ref{tab:2} compares the Orion Nebula Cluster (ONC) and W3 Main and indicates that the star formation in W3~Main shares similar molecular column density and age to the ONC (for a review of the ONC, see \citealp{2008hsf1.book..483M} and references therein).

In the ONC, \cite{Fukui_2018_Orion} identified two clouds with velocity 
difference of $\sim$4~$\kms$ which are shown to have complementary distribution 
which is consistent with a CCC that triggered the OB stars in the ONC. 
These authors presented a scenario that the ONC consists of two populations 
one was formed by the CCC in the last 0.1~Myr and the other low mass members 
, which have been continuously formed over the last 1.5~Myr. SFE in W3 Main was estimated to be $\sim$25$\%$, which is consistent with an SFE of $\sim$20$\%$ in the ONC (e.g., \citealp{Fukui_rev}). \cite{Fukui_core_mass} showed that SFE is not 
particularly enhanced in a compressed layer of CCC, and a few~$\%$ to a few times 
10$\%$ are typical values. This suggestion is also consistent with the present 
result when we consider the larger stellar ages.

The present study revealed that W3 shows evidence for triggered high-mass star 
formation by CCCs, where sufficiently high column density gas with $\sim$10$^{23}$~cm$^{-2}$ 
is colliding. \cite{Enokiya_2021_compile} compiled $\sim$50 CCC candidates and made a 
statistical study, and have shown that the number of O and early B stars are 
correlated with the molecular column density; i.e., the formation of a single 
O star requires 10$^{22}$~cm$^{-2}$ and more than ten O stars requires 10$^{23}$~cm$^{-2}$. 
W3 Main has column density of $\sim$10$^{23}$~cm$^{-2}$, and the formation of more than ten O stars is consistent with the results of \cite{Enokiya_2021_compile}. 

The molecular column density in W3(OH) is derived to be $4.1\times10^{22}$~cm$^{-2}$. 
On the other hand, the number of OB stars formed by the CCC in W3(OH) 
is 1, smaller than that in W3 Main. Nonetheless, the H$_2$O masers are associated 
with two cloud cores of 10 $\msun$ \citep{Ahmadi_2018} and they are potential 
OB stars. This is consistent with the compilation by \cite{Enokiya_2021_compile}
that the column density in W3(OH) is not enough to form ten OB stars.

W3 is an exceptionally active star formation site, while it is located in the outer solar circle where a marked decrease in the high-mass star formation rate is observed (e.g., \citealt{Djordjevic2019}). According to \citet{Fukui_rev}, among $\sim$50 Galactic CCC candidates, the objects in the outer solar circle are limited to less than 10. This prevents us from concluding that CCCs are common phenomena in the outer solar circle. However, in the recent few years, attempts to search for CCC objects in the outer solar circle have been made, and the number of CCCs is increasing (e.g., Sh2-233: \citealp{Yamada2022}, Sh2-235: \cite{Dewangan2017}, W5: \citealt{Issac2024}, AFGL 333: \citealp{Liang2021}, IRAS~01123+6430: \citealp{Koide2019}). The present work evidenced further the CCCs in the active star formation site W3, lending support for the importance of CCCs not only in the inner solar circle but also in the outer solar circle.

\begin{table*}
	\caption{Comparison of the ONC and the W3 Main cluster}
	\label{tab:2}\centering
	\begin{tabular}{lccr} 
		Parameters & ONC & W3~Main\\
    \hline
    \hline
		Total molecular column density [cm$^{-2}$] & $2\times10^{23}$ & $1.12\times10^{23}$\\
		Stellar density [pc$^{-3}$] & $1\times10^{4}$ & $4\times10^{4}$\\
        Total Cluster mass [$\msun$] & $2000$ & $4000$\\
        Age of YSOs [Myr] & $0.1$--$1$ & $0.1$--$2$ \\
        Projected velocity separation [$\kms$] & $\sim$4 & $\sim$4\\
		\hline
  \end{tabular}
\end{table*}

%% file: conc.tex
\section{Conclusions}\label{sec:conclusion}
We analyzed the $\twelvecoh$ and $\thirteencoh$ data of the W3 region \citep{Bieging2011}.  
It has been discussed as a possible scenario for three decades that 
the stars in the W3 Main and W3(OH) region have been formed by the feedback 
effect of the $\htwo$ region driven by IC~1795 (e.g., \citealp{Oey_2005}). 
It was generally thought that, on a large scale of 10 pc or more, the $\htwo$ 
region W4 compressed the gas in the HDL and triggered star formation in the HDL. 
The present data cover a smaller area than the whole W4 and W3 region and is 
not suited to test the star formation on a 20~pc scale. Instead, 
the present results have a reasonable resolution to resolve details relevant 
to pc-scale star formation in W3, where we made a detailed kinematic study and 
obtained the following results; 

\begin{itemize}
  \item The maximum gas column densities of the W3~Main and W3(OH) regions are 
  $1.1\times10^{23}$~cm$^{-2}$ and $4.1\times10^{22}$~cm$^{-2}$, respectively. 
  The first moment distribution revealed that the W3 region consists of at least three clouds. The $-43$~$\kms$ cloud is the most extended, and it is distributed in both the W3 Main and W3(OH) regions. The $-39$~$\kms$ component is localized in W3 Main region, while the $-50$~$\kms$ cloud overlaps the $-43$~$\kms$ cloud in W3(OH). The total gas mass of the region analysed by the present study amounts to 21000~$\msun$, while three clouds have masses of 4900~$\msun$ ($-39$~$\kms$ cloud), 6800~$\msun$ ($-43$~$\kms$ cloud), and 3000~$\msun$ ($-50$~$\kms$ cloud). 

  \item In W3~Main, we tested if the feedback is important in the cloud kinematics by 
  comparing the stellar feedback energy with the kinetic energy of the clouds. 
  The most energetic source near W3~Main is IC~1795, which includes ten high-mass 
  stars and the $\htwo$ region. We estimated that the total kinetic energy of these 
  stars/$\htwo$ region is not large enough to affect the gas kinematics of the two clouds, 
  nor does the local velocity distribution at the boundary of the $\htwo$ region show no 
  velocity shift. We suggest that any feedback by the $\htwo$ region or IC~1795 or the 
  molecular outflow is not significant in altering the cloud morphology or kinematics. 
  Accordingly, we suggest that stellar feedback does not play a role in W3.

  \item In W3 Main, the $-39$~$\kms$ cloud has a diameter of $\sim$5~pc, while the $-43$~$\kms$ cloud is extended along the Galactic plane by more than 25~pc with a ``cavity'' of $\sim$5~pc diameter with significantly decreased CO intensity in the eastern edge. We find that the two clouds show spatially complementary distribution with a displacement of 4.1 pc, where the $-39$~$\kms$ cloud fits the cavity of the $-43$~$\kms$ cloud. We also find that the clouds show a V-shape in a position-velocity diagram. These signatures are consistent with the fact that the two clouds are colliding with each other to deform their original distribution and kinematics into a merging cloud. The displacement indicates that the $-39$~$\kms$ cloud has moved to the northwestern direction in the past. The location of W3 Main, which is in the northwest of the cavity, is consistent with W3 Main, where ten OB stars are localised and correspond to the collision-compressed layer as modelled by \citet{Habe_Ohta}. 

  \item The collision in W3 Main explains the formation of the most massive star IRS 2 and 10 OB stars. The W3 Main region also includes several outflows, indicating star formation activity with a timescale of $10^4$ yr. This short time scale is consistent with the scenario that the W3 Main is now at the collision-compressed layer, and currently triggered star formation is ongoing. On the other hand, we estimate the typical timescale of the collision to be 1--2 Myr by dividing the displacement 4.1 pc by the relative velocity of the clouds $\sim$4~$\kms$, depending on the assumption of the angle between collision direction and line-of-sight. This collision to form the cavity toward IC~1795 may explain the formation of the IC~1795 cluster because the age of the IC~1795 may be as young as 2 Myr, which is consistent with the stellar age.
  
  \item The $-50$ and $-43$~$\kms$ clouds in W3(OH) are also complementary in the spatial distribution in the sense that the $-50$~$\kms$ cloud is located on the east of the $-43$~$\kms$ cloud on a scale of 3--5~pc. The spatial distribution seems to be simpler than in W3~Main, and the total cloud mass involved is less than a factor of 2 than the 
  W3~Main cloud.  We suggest a possible scenario in which the two clouds are colliding 
  with each other, where the time scale of the collision is roughly estimated to 
  be 0.7~Myr, shorter than in W3~Main. We conservatively note that the evidence 
  for the collision is not as strong as in W3~Main, for which the complicated CO 
  distribution lends firmer support for the complementary distribution typical to 
  a CCC.

  \item W3 is experiencing a CCC in two places now. In W3~Main, the maximum column 
  density is $1.1\times10^{23}$~cm$^{-2}$ for a projected velocity difference of $\sim$4~$\kms$ and more than ten O stars are formed. These regions are the sites of high-mass star formation 
  relatively close to the sun. The physical parameters are similar to the ONC. W3(OH) 
  has a somewhat smaller maximum column density for the projected velocity difference of 
  $\sim$4~$\kms$ and a few high mass star candidates for O star are formed. These are well 
  consistent with the 50 samples of CCC candidates \citep{Enokiya_2021_compile,Fukui_rev}. 

\end{itemize}

The W3 region has attracted keen interests in the last few decades, 
whereas there has been no unified picture of star formation, including 
the effects of triggering. This would be due to the paucity of investigations 
of molecular gas, where the most direct and recent traces of star formation 
are carved. We, therefore, studied the CO gas in detail. As a result, clear signatures 
of CCCs have been revealed. Our model frames a scenario which offers a 
unified view of the star formation over a few Myrs. The method naturally has no 
direct explanation for the cluster, which has no associated molecular gas, whereas 
it is still possible that the cluster was formed by a past event whose relic 
has already been dispersed. A stellar system with age spread and better statistics is essential.

%% file: ack.tex
\begin{ack}
We are grateful to Professor Philippe Andr{\'e} for kindly providing {\it Herschel} data. We also acknowledge Akio Taniguchi, Keisuke Sakasai, Kazuki Shiotani for their valuable support during data analysis. The Heinrich Hertz Submillimeter Telescope is operated by the Arizona Radio Observatory, a part of Steward Observatory at The University of Arizona. PACS has been developed by a consortium of institutes led by MPE (Germany) and including UVIE (Austria); KU Leuven, CSL, IMEC (Belgium); CEA, LAM (France); MPIA (Germany); INAF-IFSI/OAA/OAP/OAT, LENS, SISSA (Italy); IAC (Spain). This development has been supported by the funding agencies BMVIT (Austria), ESA-PRODEX (Belgium), CEA/CNES (France), DLR (Germany), ASI/INAF (Italy), and CICYT/MCYT (Spain). SPIRE has been developed by a consortium of institutes led by Cardiff University (UK) and including Univ. Lethbridge (Canada); NAOC (China); CEA, LAM (France); IFSI, Univ. Padua (Italy); IAC (Spain); Stockholm Observatory (Sweden); Imperial College London, RAL, UCL-MSSL, UKATC, Univ. Sussex (UK); and Caltech, JPL, NHSC, Univ. Colorado (USA). This development has been supported by national funding agencies: CSA (Canada); NAOC (China); CEA, CNES, CNRS (France); ASI (Italy); MCINN (Spain); SNSB (Sweden); STFC, UKSA (UK); and NASA (USA). This research made use of Astropy,\footnote{http://www.astropy.org} a community-developed core Python package for Astronomy \citep{astropy:2013, astropy:2018}. This research made use of APLpy, an open-source plotting package for Python hosted at http://aplpy.github.com. This work was supported by JSPS KAKENHI Grant Numbers JP15H05694, JP18K13580, JP19K14758, JP19H05075, JP20K14520, JP20H01945, and 22KJ1604. R.Y. is a Research Fellow of JSPS. 
\end{ack}

%% file: appendix.tex
\appendix
\section{Calculations of column densities and masses}\label{A1}
\begin{figure*}
	\includegraphics[width=17cm]{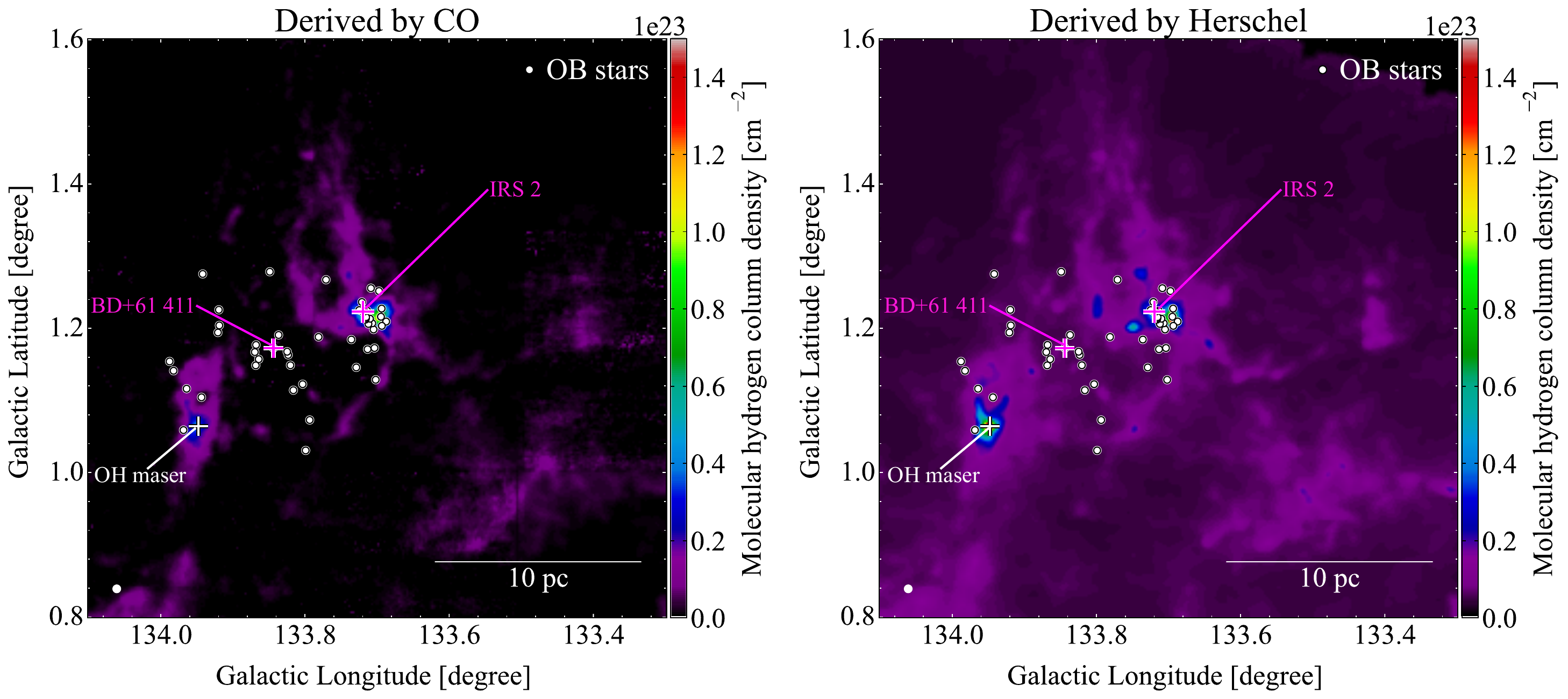}
    \caption{Colum density maps derived from CO (a) and Herschel (b) respectively (This study and \citealt{Rivera2013}). The magenta crosses, white cross, and white dots indicate the positions of BD+61 411, IRS~2, a driving source of the OH maser, and well-known OB stars (e.g., \citealt{Navarete2019}). }
    \label{fig15}
\end{figure*}

\begin{figure*}
	\includegraphics[width=17cm]{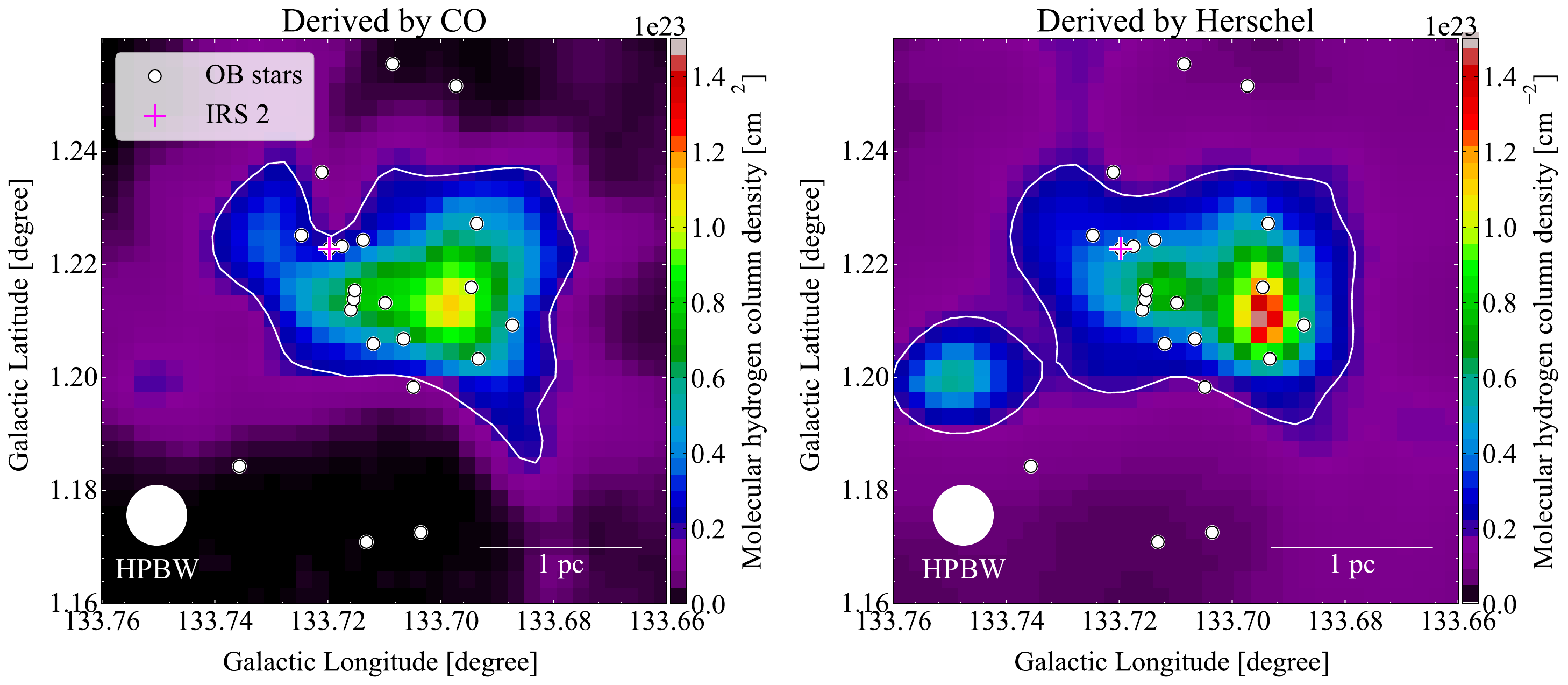}
    \caption{Close up of the W3 Main region of Figure \ref{fig16}. The magenta crosses, white cross, and white dots indicate the positions of IRS2 and well-known OB stars (e.g., \citealt{Navarete2019}).}
    \label{fig16}
\end{figure*}

We assume the local thermodynamic equilibrium (LTE) and calculate the molecular column density and mass of each cloud. We used only the pixels having physical parameters of more than 6$\sigma$. By assuming that the $\twelvecoh$ emission is optically thick, the excitation temperature $T_\mathrm{ex}$ is given as follows;
\begin{equation}
    T_{\mathrm{ex}}=11.06\left\{\ln\left[1+\frac{11.06}{T_{\mathrm{peak}}+0.19}\right]\right\}^{-1}.
\end{equation}
From this equation, we derived $T_{\mathrm{ex}}$ for every pixel. The equivalent brightness temperature $J(T)$ is expressed as below for Planck constant $h$, Boltzmann constant $k_\mathrm{B}$, and the observed frequency $\nu$,
\begin{equation}
    J(T)=\frac{h\nu}{k_\mathrm{B}}\left[\exp{\left(\frac{h\nu}{k_{\mathrm{B}}T}\right)}-1\right]^{-1}.
\end{equation}
The radiation transfer equation gives the $\thirteencoh$ optical depth $\tau(\nu)$ to be, 
\begin{equation}
    \tau(v)=-\ln{\left[1-\frac{T_{\mathrm{mb}}}{J(T_{\mathrm{ex}})-J(T_{\mathrm{bg}})}\right]}.
\end{equation}
And the $\thirteencoh$ column density as follows; 
\begin{equation}
N=\sum_v \tau(v)\Delta v \frac{3k_\mathrm{B}T_{\mathrm{ex}}}{4\pi^3\nu\mu^2}\exp{\left(\frac{h\nu J}{2k_{\mathrm{B}}T_{\mathrm{ex}}}\right)}\times \frac{1}{1-\exp{(-h\nu/k_{\mathrm{B}}T_{\mathrm{ex}})}},
\end{equation}
where we used $k_\mathrm{B}$ = $1.38 \times 10^{-16}$ (erg K$^{-1}$), $\nu = 2.20\times10^{11}$~(Hz), 
$\mu = 1.10 \times 10^{-19}$~(esu cm), $h = 6.63 \times 10^{-27}$~(erg s), $J = 1$, and $T_\mathrm{ex}$ was adopted from each pixel. By assuming that the ratio $N_\mathrm{H_2}/N_{^{13}\mathrm{CO}}$ to be $7.1\times10^5$ \citep{Frerking}, the column density is calculated. The molecular mass is given by
\begin{equation}
M = m_{\mathrm{p}} \mu_\mathrm{m} D^2\Omega \sum_{i}N_{i}(\mathrm{H_2}), 
\label{mass}
\end{equation}
where $\mu_m$, $m_\mathrm{p}$, $D$, $\Omega$,  $N_{i}(\mathrm{H_2})$ are the mean molecular weight, the proton mass, distance, solid angle, and column density of the $i$-th pixel. If the helium abundance is assumed to be 20$\%$, $\mu_m$ is 2.8. The distances of W3 Main and W3(OH) are 2.00~kpc (e.g., \citealp{Navarete2019}). The derived parameters are listed in Table~\ref{tab:1}. 

\section{Comparison of CO-derived and dust-derived molecular hydrogen column dnesity}
In this study, we utilize the physical parameters of molecular clouds derived from CO using the procedure outlined in Appendix \ref{A1}. The column densities derived from CO and \textit{Herschel} data are consistent. Figures \ref{fig15}a and \ref{fig15}b depict maps of hydrogen column density obtained from \textit{Herschel} and CO, respectively. These figures are displayed with aligned colour scales, revealing a similarity in spatial distribution between the two datasets. Figure \ref{fig16} presents an enlarged view of the W3 Main. While regions with high hydrogen column densities generally exhibit optical thickness in $^{13}$CO, the differences with column densities derived from \textit{Herschel} are within a factor of approximately 2, suggesting that the impact of optical thickness is not significant.